\documentclass[manuscript]{aastex}

\slugcomment{Submitted to AJ}

\shorttitle{Near-IR WR Survey}
\shortauthors{Shara et al}

\begin{document}

\title{A Near-Infrared Survey of the Inner Galactic Plane for Wolf-Rayet Stars I. Methods and First Results: 41 New WR Stars}

\author{Michael M. Shara}
\affil{American Museum of Natural History, 79th Street and Central Park West, New York, NY, 10024-5192}
\email{mshara@amnh.org}

\author{Anthony F. J. Moffat}
\affil{D\'epartement de Physique, Universit\'e de Montr\'eal, CP 6128 Succ. C-V, Montr\'eal, QC, H3C 3J7, Canada }
\email{moffat@astro.umontreal.ca }

\author{Jill Gerke}
\affil{American Museum of Natural History, 79th Street and Central Park West, New York, NY, 10024-5192}
\email{jgerke@amnh.org}

\author{David Zurek}
\affil{American Museum of Natural History, 79th Street and Central Park West, New York, NY, 10024-5192}
\email{dzurek@amnh.org}

\author{Kathryn Stanonik}
\affil{Department of Astronomy, Columbia University, 550 West 120th Street, New York, NY 10027}
\email{keejo@astro.columbia.edu }

\author{Ren\'e Doyon }
\affil{D\'epartement de Physique, Universit\'e de Montr\'eal, CP 6128, Succ. C-V, Montr\'eal, QC, H3C 3J7, Canada }
\email{doyon@astro.umontreal.ca }

\author{Etienne Artigau}
\affil{Gemini Observatory, AURA,
Casilla 603, La Serena, Chile}
\email{eartigau@gemini.edu}

\author{Laurent Drissen }
\affil{D\'epartement de Physique, Universit\'e Laval, Pavillon Vachon, Quebec City, QC, G1K 7P4 Canada}
\email{ldrissen@phy.ulaval.ca}

\author{Alfredo Villar-Sbaffi }
\affil{ D\'epartement de Physique, Universit\'e de Montr\'eal, CP 6128, Succ. C-V, Montr\'eal, QC, H3C 3J7, Canada }
\email{alfredovs@hotmail.com}

\begin{abstract}
The discovery of new Wolf-Rayet (WR) stars in our Galaxy via large-scale narrowband optical surveys has been severely limited by dust extinction. Recent improvements in infrared technology have made narrowband-broadband imaging surveys viable again. We report a new J, K and narrow-band imaging survey of 300 square degrees of the plane of the Galaxy, spanning 150 degrees in Galactic longitude and reaching 1 degree above and below the Galactic plane. The survey has a useful limiting magnitude of K = 15 over most of the observed Galactic plane, and K = 14 within a few degrees of the Galactic center. Thousands of emission line candidates have been detected. In spectrographic follow-ups of 173 WR star candidates we have discovered 41 new WR stars, 15 of type WN and 26 of type WC. Star subtype assignments have been confirmed with K band spectra, and distances approximated using the method of spectroscopic parallax. A few of the new WR stars are amongst the most distant known in our Galaxy. The distribution of these new WR stars is seen to follow that of previously known WR stars along the spiral arms of the Galaxy. Tentative radial velocities were also measured for most of the new WR stars.

\end{abstract}


\section{Introduction}

\subsection{Motivation}

It is extraordinary but true that the galaxy of the Local Group whose global stellar populations are least well observed is our own Milky Way. Deeply immersed within the optically opaque, dusty lanes of our Galaxy's spiral arms, astronomers have been frustrated for centuries in their attempts to map the stellar populations of the Milky Way. A complete census of our Galaxy for every member of even one class of star would have seemed like an impossible goal even a decade ago. Major advances in instrumentation are transforming this daunting task from near-impossibility to increasingly likely. Wide-field, high-resolution, sensitive surveys, particularly in the near-infrared and X-ray parts of the electromagnetic spectrum (where the Galaxy is relatively transparent), are key to locating and characterizing all members of one or more classes of stellar object. The goal of the project described in this paper is to detect and spectrographically characterize at least 90\% of the Wolf-Rayet stars in the Milky Way within ten years.

A set of well-defined tests of stellar evolution theory will follow from detections of complete samples of Wolf-Rayet and other related stars. For example, the radial abundance gradient across our Galaxy and the increase of the WR/O number-ratio with increasing Z suggests that many more WR stars will be found in the inner parts of the Milky Way than in the outer regions. In addition, our previous HST survey of the HII regions in the ScIII galaxy NGC 2403 \citep{dri99} suggested that the distribution of Wolf-Rayet and red supergiant stars (RSG) is a sensitive diagnostic of the recent star-forming history of these large complexes: young cores of O and WR stars are surrounded by older halos containing RSG. Theory predicts that the number-ratio WR/O increases with increasing metallicity; thus, relatively fewer WR stars form at lower Z. We will also be able to determine if superclusters, dominated by Wolf-Rayet stars, are common in the Milky Way. Finally, we note that WR stars are predicted to end their lives as supernovae, and in rare cases as Gamma Ray Bursts. The Wolf-Rayet stars in the Milky Way may be abundant enough for one to erupt as a Type Ib or Ic supernova within a few generations.  This comes from the assumption that the MW contains $\sim$6000 WR stars, each lasting $\sim5\times10^5$ yrs. The clear identification of a WR star as the progenitor of one of these eruptions would be a dramatic confirmation of a key prediction of stellar evolution theory.

\subsection{Wolf-Rayet stars}

Wolf-Rayet (WR) stars are massive (with initial masses greater than $\sim 20 M_\sun$ at $Z_{\odot}$) stars with strong winds ($\dot{M} \sim 10^{-5} M_\sun yr^{-1}$ ) displaying the heavier elements created by what are normally internal nuclear processes. Distinctive spectra with strong, broad emission lines of helium, and either nitrogen (WN) or carbon (WC) are the defining observational characteristics of WR stars. As they have relatively short lifetimes (about $5\times10^5$ years), WR stars are excellent tracers of star formation, and they are also believed to be type Ib or Ic supernova progenitors, because they have removed their outer H-rich layers (WN) or even He-rich layers (WC/O).

About 300 WR stars have been previously identified in the Milky Way \citep{vdh06}, with distribution models predicting $\sim$1000-6500 total expected (\cite{sha99}; \cite{vdh01}). Optical narrow band surveys have been severely limited by interstellar extinction \citep{sha99}, and so the natural solution is to turn to the near infrared. Emission line magnitudes of 40 known WR stars are presented in Appendix A. A model of the Milky Way, predicting the numbers and distributions of WR stars visible in the K band is presented in Appendix B.

In an initial attempt at a narrowband near-infrared survey \cite{hom03} had limited success, while \cite{had07} were somewhat more successful with their color-based selection of objects from the 2MASS \citep{skr06} and GLIMPSE \citep{ben03} surveys.

Utilizing a new narrow-band survey, described in $\S$ \ref{observations} along with the candidate selection methods, we have found 41 new WR stars: 15 WN and 26 WC. Spectrographic follow up and data reduction is described in $\S$ \ref{specobsred}. Our resulting K-band spectra, line measurements, and subtype classification are presented in $\S$ \ref{results}. Distances to the new WR stars are calculated and their distribution within the Galaxy considered in $\S$ \ref{distribution}. Our conclusions are summarized in $\S$ \ref{conclusion}.

\section{Observations} \label{observations}

The most reliable optical technique to detect individual Wolf-Rayet stars in crowded fields consists of subtracting a normalized continuum image from an image obtained with a narrowband filter centered on the HeII 4686\AA ~line. This works well for individual WR stars with equivalent widths of HeII(4686) $10-15$ \AA ~or larger, and even for dense, unresolved clusters that include a very small fraction of WR stars (\cite{dri93}). Unfortunately, dust extinction makes this technique infeasible for the large majority of Galactic WR stars. Only in the near-infrared can we hope to detect the WR stars farther than about 5 kpc.

A near-infrared survey of the plane of the Galaxy was carried out under the umbrella of the SMARTS consortium (see http://www.astro.yale.edu/smarts/). The imaging data were taken over approximately 200 nights in 2005-2006 on the Cerro Tololo Inter-American Observatory (CTIO) 1.5 meter telescope, using the Universit\'e de Montr\'eal's CPAPIR camera. Images are 35$^\prime$ on a side, with a plate scale of 1.03$^{\prime\prime}$ per pixel, and cover  1$^\circ$ above and below the Galactic plane from Galactic longitude \textit{l}=-90$^\circ$ to \textit{l}=60$^\circ$. Each of the 1200 fields was imaged in the J and K bands, and through a selection of narrow bands designed for identifying WR stars.

Motivated by the WR infrared spectra in Figer et al. (1997), we purchased a filter set (Table \ref{filters}) which targeted the emission line features He \textsc{i} 2.062 $\mu$m, C \textsc{iv} 2.081 $\mu$m, H \textsc{i} Br$\gamma$ 2.169 $\mu$m, and He \textsc{ii} 2.192 $\mu$m. In addition, two narrow-band continuum filters were used which were selected to be relatively devoid of emission lines, one blueward (2.033 $\mu$m) and one redward (2.255 $\mu$m) of the emission line filters. These were then used to linearly interpolate a continuum magnitude at each of the emission line bands, so we could calculate the difference in measured and interpolated magnitudes, $\Delta m$, indicative of emission or absorption in that band. In this paper, we are working from a catalogue of the calculated $\Delta m$ values for all of the stars contained in the survey area which had been processed by the time of our spectroscopic follow-up. This consisted of about 75\% of the survey area, and in general we excluded those areas most crowded with stars, in particular the approximately eight degrees in longitude closest to the Galactic center.

\subsection{Image Processing}

The more than $77,000$ science and dome flat images of the survey require a customized, streamlined pipeline to reduce this large amount of data. The pipeline was constructed (by JG and DZ) in IDL and uses the 2MASS catalogue extensively as a reference for both astrometry and photometry. Each of the $\sim1200$ fields has been imaged in each of the $8$ filters at $7$ dither positions with separations of $\sim 15\arcsec$. Dome flats were used to flatten each of the images for sensitivity and chip defects. These $7$ dither positions were median combined together without shifting to remove most of the stellar sources and to create a sky image. The sky image was subtracted from each image.

Sources for each field/filter were matched to 2MASS to determine the world coordinate system (WCS). Once a WCS was fit, another iterative process was performed to minimize the residuals to 2MASS and determine the best geometric distortion solution. The IDL procedure WARP\_TRI applied the geometric distortion solution. The IDL procedure HASTOM aligned all the images taken in a dither series. These images were combined to create a final deep exposure.

\subsection{Photometry}

Sources were identified as WR candidates through emission in the narrowband filters.  This made it necessary to have not only the magnitude of a source in the narrowband filter, but also the magnitude of the continuum at that wavelength.  As a result, each narrowband filter had to be examined concurrently with the CONT1 and CONT2 filters.

Sources on the final deep exposures were detected using the IDL procedure FIND. Aperture photometry was then carried out using the IDL procedure APER with a 2-pixel aperture and a sky annulus from 10 to 20 pixels. Objects matched to the 2MASS catalog (at least 100 in each field) were used to determine a zero point for each filter in each deep image. An object with a flat spectrum through our filters has the same magnitude in all 6 filters.

Sources were considered matched across filters if their positions were consistent within 0.5 pixel. These matched sources with 2MASS-calibrated magnitudes were then used to construct emission-magnitude diagrams (EMDs). An EMD was constructed for each narrowband filter with the wavelength-interpolated continuum magnitude vs continuum-subtracted narrowband magnitude. The stars scatter around a continuum-subtracted narrowband magnitude of zero. In each magnitude bin we calculate the standard deviation sigma of the continuum-subtracted narrowband magnitude. Objects that are 5$\sigma$ or more in the negative direction from the locus of stars are considered candidates.  
We determined the offset to convert instrumental magnitudes to apparent magnitudes using the 2MASS Ks band catalogue.  The images were divided into an 8x8 grid, with each of the 64 areas having an individually determined offset, to compensate for an observed color dependence (probably due to variable reddening) across the field. The IDL procedure APER determined the magnitude for the source at the coordinates given by 2MASS.

Once the offsets to convert to apparent magnitude were calculated, FIND, an IDL procedure, identified sources in an image that were a given deviation above the background.  APER found the instrumental magnitude for the sources, which were then converted to apparent magnitudes.  The CONT1 filter was taken as a reference image and HASTROM aligned all filters from a field.  The sources from CONT1 were then matched to the sources from the CONT2 image and sources within 0.5 pixels were kept as matches.  The matched sources were then compared to the sources of a narrowband filter.  The result was a list of sources found within 0.5 pixels of each other in both of the continuum filters and the narrowband filter.  A linear fit was found between the CONT1 and CONT2 filter magnitudes of each source. Then, using the central wavelength for the narrowband filter, an interpolated continuum value at the narrowband wavelength was determined.  The magnitude of the source in the narrowband filter was then subtracted from the interpolated continuum value, giving the negative emission magnitude for the source in that narrowband filter. The emission magnitude for a source was also estimated by subtracting the CONT1 and the CONT2 magnitude from the narrowband magnitude, resulting in a total of three estimates for the emission magnitude of a source.  
An EMD of continuum magnitude vs. emission magnitude was created for each narrowband filter and the standard deviation was determined for the sources in bins of 1 magnitude.  Sources that had emission magnitudes of $5\sigma$ or greater from the center of the EMD were marked as WR candidates.  1 arcmin $\times$ 1 arcmin finder charts were produced for each WR candidate, showing the candidate in all filters of the survey and also in XDSS red and XDSS infrared images. These are presented in Appendix C.  Candidates with emission magnitudes that were similar to those of the known WR stars covered in the survey were selected for spectrographic follow up.  This refined list of candidates was blinked by eye to remove any candidates that did not resemble stars.

\subsection{Candidate Selection}

In this initial, exploratory phase of the survey we used two techniques when selecting targets for spectroscopic follow-up. We began by selecting targets with such powerful emission lines that the star appeared brighter in the narrow-band images when compared to the continuum images even when examined by eye. This corresponds to a minimum 0.5 to 1 magnitude difference in brightness between the narrow- and broad-band images. We initially selected candidates displaying a brightening of at least 0.5 magnitudes in at least one narrow band image relative to the continuum. This resulted in the detection of 34 new planetary nebulae (which will be reported elsewhere) whose very strong, sharp emission lines, He \textsc{i} 2.058 $\mu$m and Br$\gamma$ 2.166 $\mu$m, fell within our He \textsc{i} and Br$\gamma$ narrow-band imaged fields. A few of the planetary nebulae were slightly resolved on the Br$\gamma$ narrow-band images. No new WR stars have yet been found this way.

Our second, much more successful, technique relied on using known WR stars to calibrate our selection of targets. Forty known WR stars were selected within the survey area, and patterns were found in their narrow band $\Delta m$ values which distinguished broad subtypes of WR stars. (See Appendix \ref{earlywork}). WC stars generally showed strong (-0.8 magnitudes or less) emission excess in the C \textsc{iv} filter, and slightly weaker emission (between -0.4 and -0.8 magnitudes) in the He \textsc{i} filter, due to the blue side of the C \textsc{iv} line extending into the range of the He \textsc{i} narrow-band filter. Early WN stars generally showed moderate emission in the Br$\gamma$ filter and the He \textsc{ii} filter, but slight absorption (0.1 magnitudes) in both He \textsc{i} and C \textsc{iv}.

Using these criteria, 173 candidate targets were selected which appeared at least $5\sigma$ brighter in their narrow-band filter than did other stars in the field and which also fit the criteria suggested by the known WRs. Though we are reliably detecting stars to magnitude 14-14.5 in all of the filters (by judicious choice of exposure times), during this exploratory stage candidates were selected to have emission-band magnitudes brighter than K = 11.5. This is because the initial spectrographic follow-up is being done with a 1.5m telescope (see below). (Thousands of uncrowded $5\sigma$ candidates as faint as faint as K = 14.5 will be the subjects of future papers). There are strong selection effects for those WR subtypes which were used to determine the selection criteria, and as a result, no WC9 or late WN type ($>$WN6) stars have yet been found. Improvements now underway in survey image reduction will permit discovery of the less strongly distinguished subtypes.

\section{Spectrographic Observations and Reduction} \label{specobsred}

The spectrographic follow-up data were taken between 28 April and 6 June 2007, with the near-infrared (0.8 - 2.5 $\mu$m) SIMON spectrograph \citep{doy00} of the Universit\'e de Montr\'eal mounted on the 1.5 meter telescope at CTIO. SIMON has a scale of 0.46 "/pixel on the CTIO 1.5 meter telescope. Targets were observed in the K band with a resolving power of R $\sim$ 1500. Each target was observed 5 times, with a nod to move the target along the slit between each observation. Total integration times ranged from 10 to 30 minutes per candidate.

All data were reduced using \textsc{iraf} routines. Images were dark subtracted and flat fielded to remove any instrument signature; spectra were then extracted using the \textsc{apall} task. This task also provided sky subtraction by fitting the background on either side of the object along the slit. The five exposures were scaled and median-combined. Standard stars, which were observed periodically throughout the night, were similarly reduced, and corrections made for Br$\gamma$ absorption in the telluric standard. Object spectra were then divided by the temporally closest standard-star spectra to remove, as best as possible, the atmospheric absorption features, particularly those at 2.008 $\mu$m and 2.059 $\mu$m, and a fainter feature at 2.199 $\mu$m.

Wavelength calibrations were done using the atmospheric OH emission observed off-target by the spectrograph during each object observation. Lines were first identified using the \textsc{identify} task and were compared with the coordinate list ohlines.dat included therein (\cite{ste79}). The wavelength solution was then refitted to each observed target using the \textsc{reidentify} task. In general, the root-mean-square error of the residuals of wavelength fits was less than 1 \AA, and in most cases between 0.1 and 0.3 \AA. These dispersion solutions were each applied to the corresponding target object, and then corrected for the intrinsic heliocentric motion using the \textsc{rvcorrect} task to identify the heliocentric velocity and the \textsc{dopcorrect} task to Doppler shift the wavelength scale.

%(num targets)(seeing)(airmass, et)
%slitwidth, dispersion 3.4 Ang/pixel
%plate scales?
%S/N?
%5 exposures each target, nodding along ~20 pixels along slit
%\clearpage

\section{Results} \label{results}
\begin{figure}[t]
\epsscale{0.9}
\figurenum{1a}
\plotone{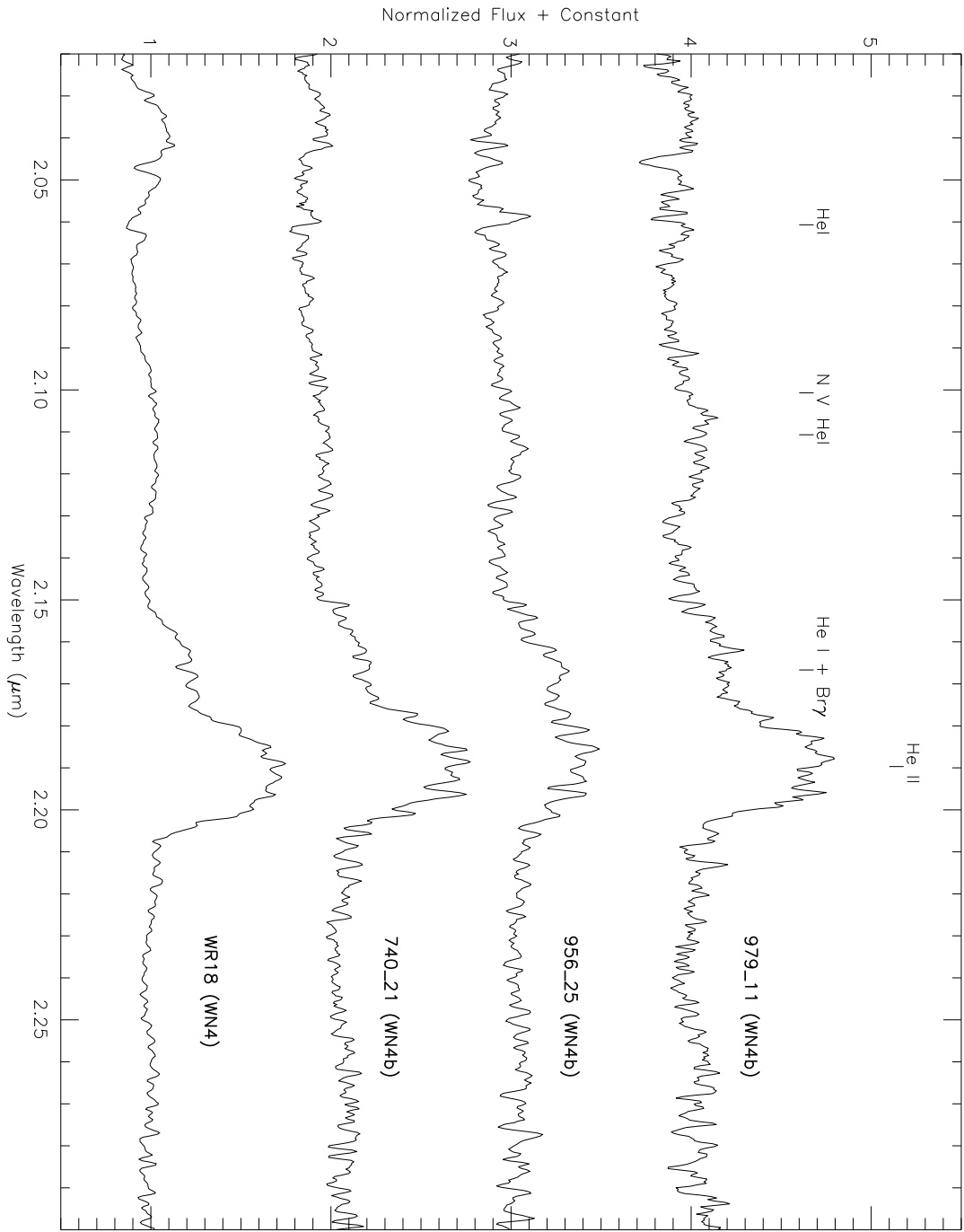}
\caption{WN4 spectra. \label{wn4}}
\end{figure}

\begin{figure}[t]
\figurenum{1b}
\plotone{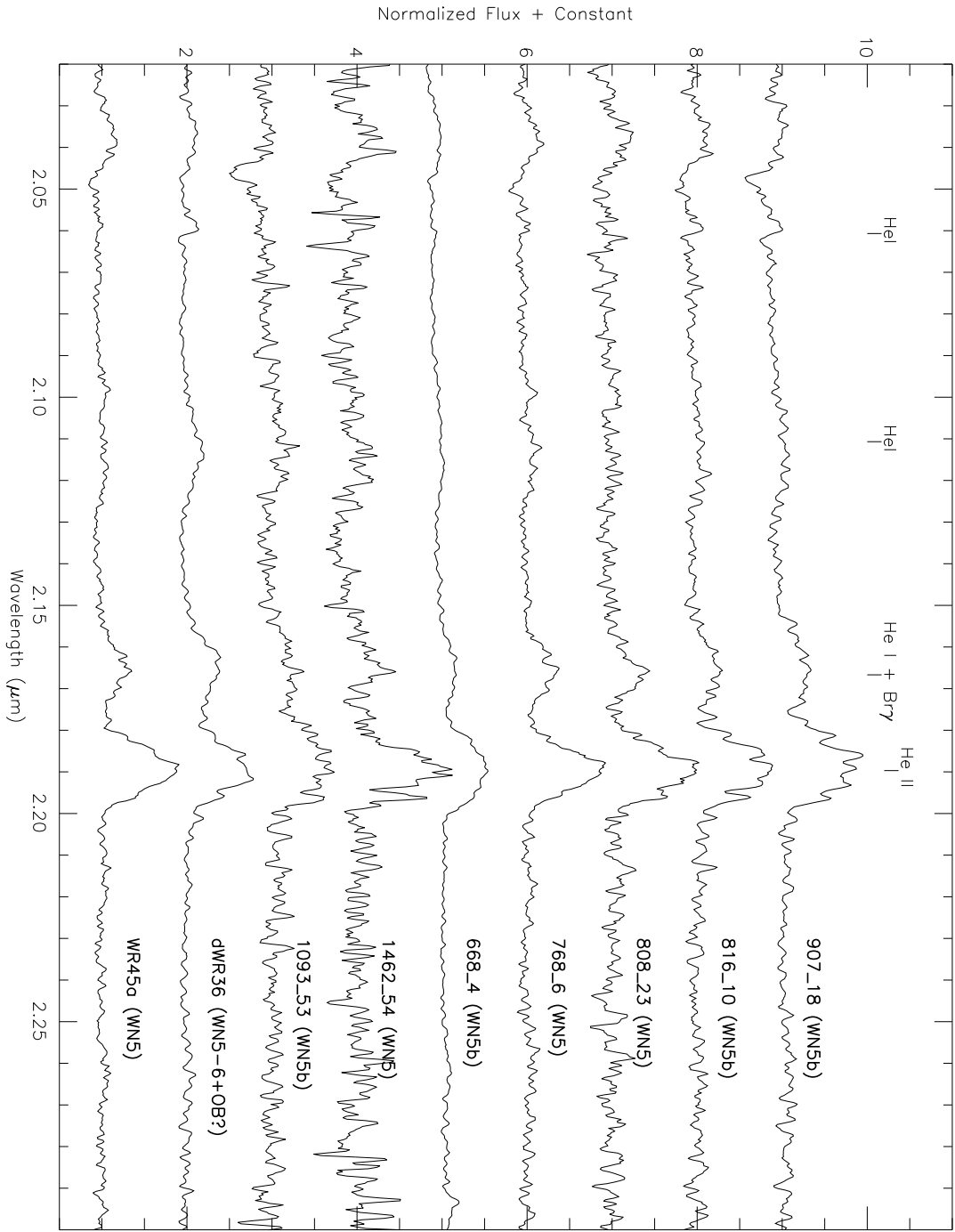}
\caption{WN5 spectra. \label{wn5}}
\end{figure}

\begin{figure}[t]
\figurenum{1c}
\plotone{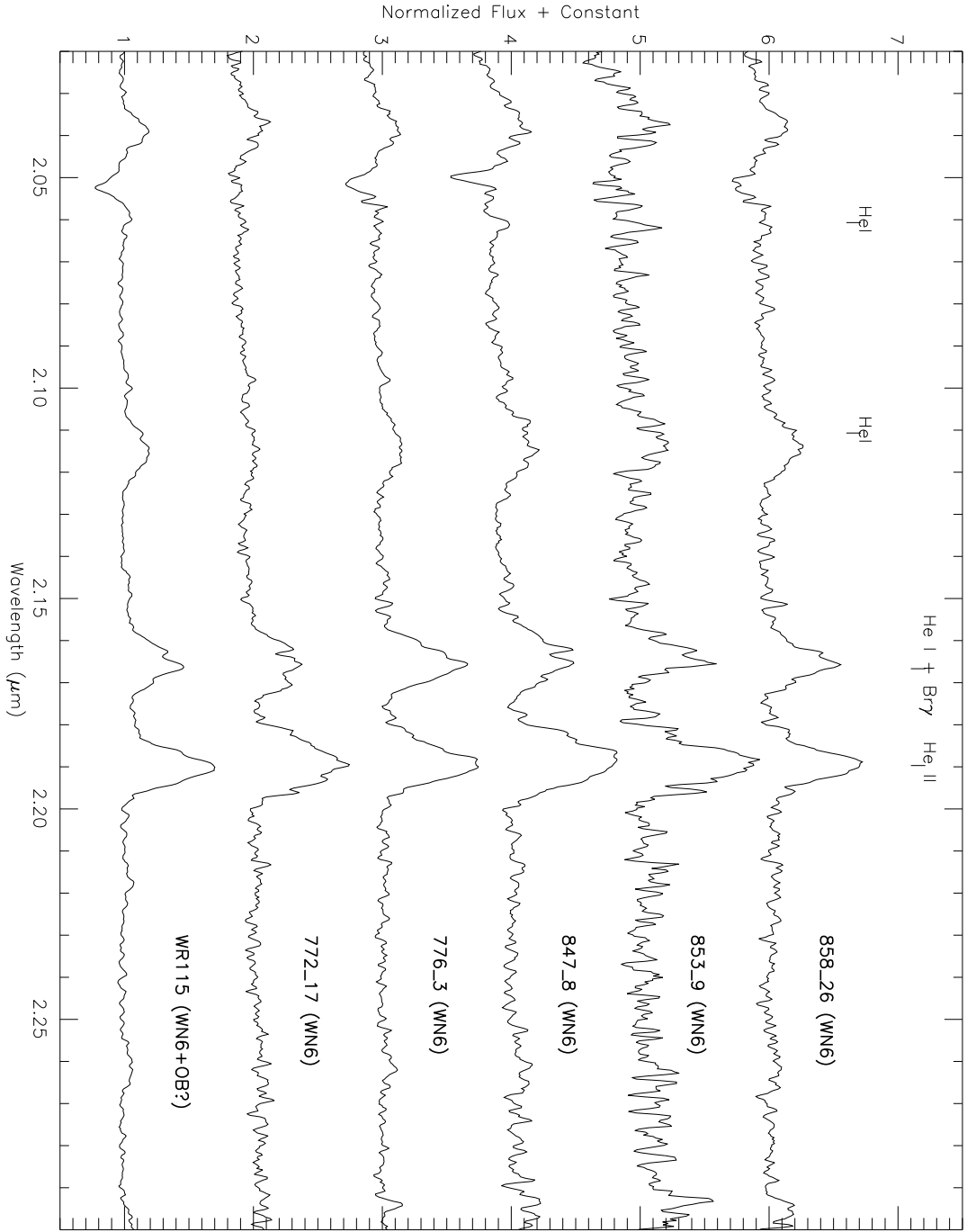}
\caption{WN6 spectra. \label{wn6}}
\end{figure}

\begin{figure}[t]
\figurenum{1d}
\plotone{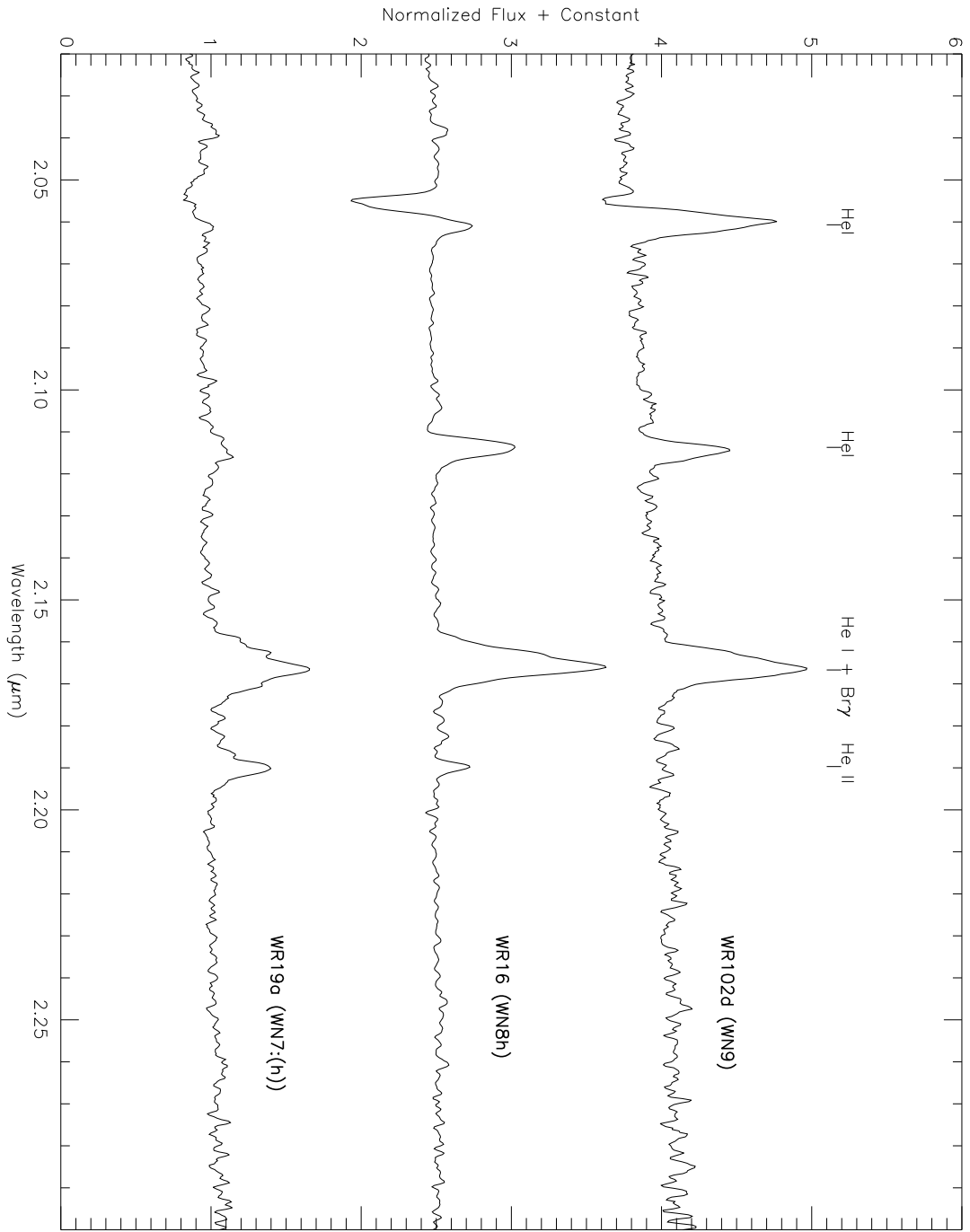}
\caption{WN7-9 spectra. \label{wnex}}
\end{figure}

\begin{figure}[t]
\figurenum{1e}
\plotone{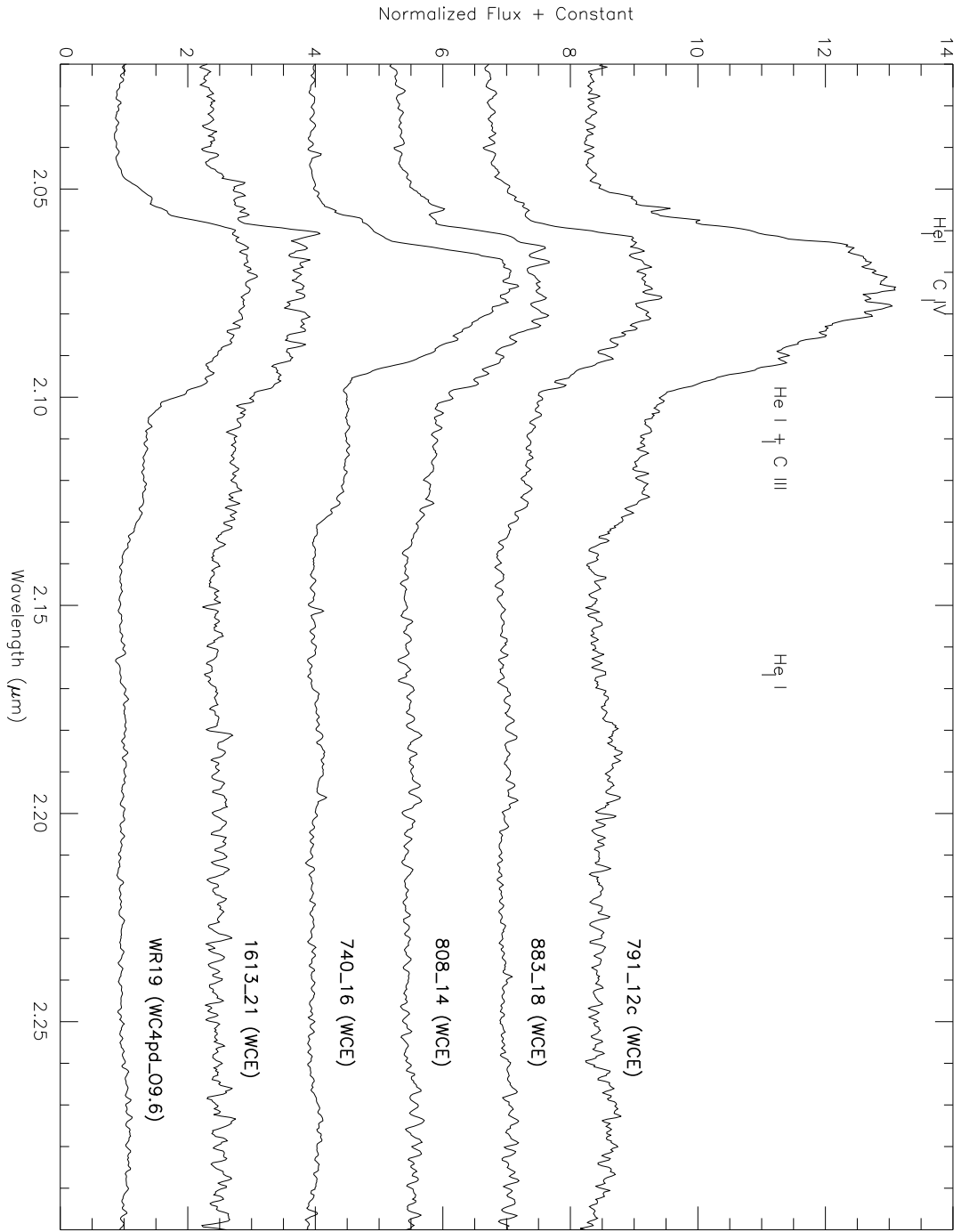}
\caption{WCE spectra. \label{wce}}
\end{figure}

\begin{figure}[t]
\figurenum{1f}
\plotone{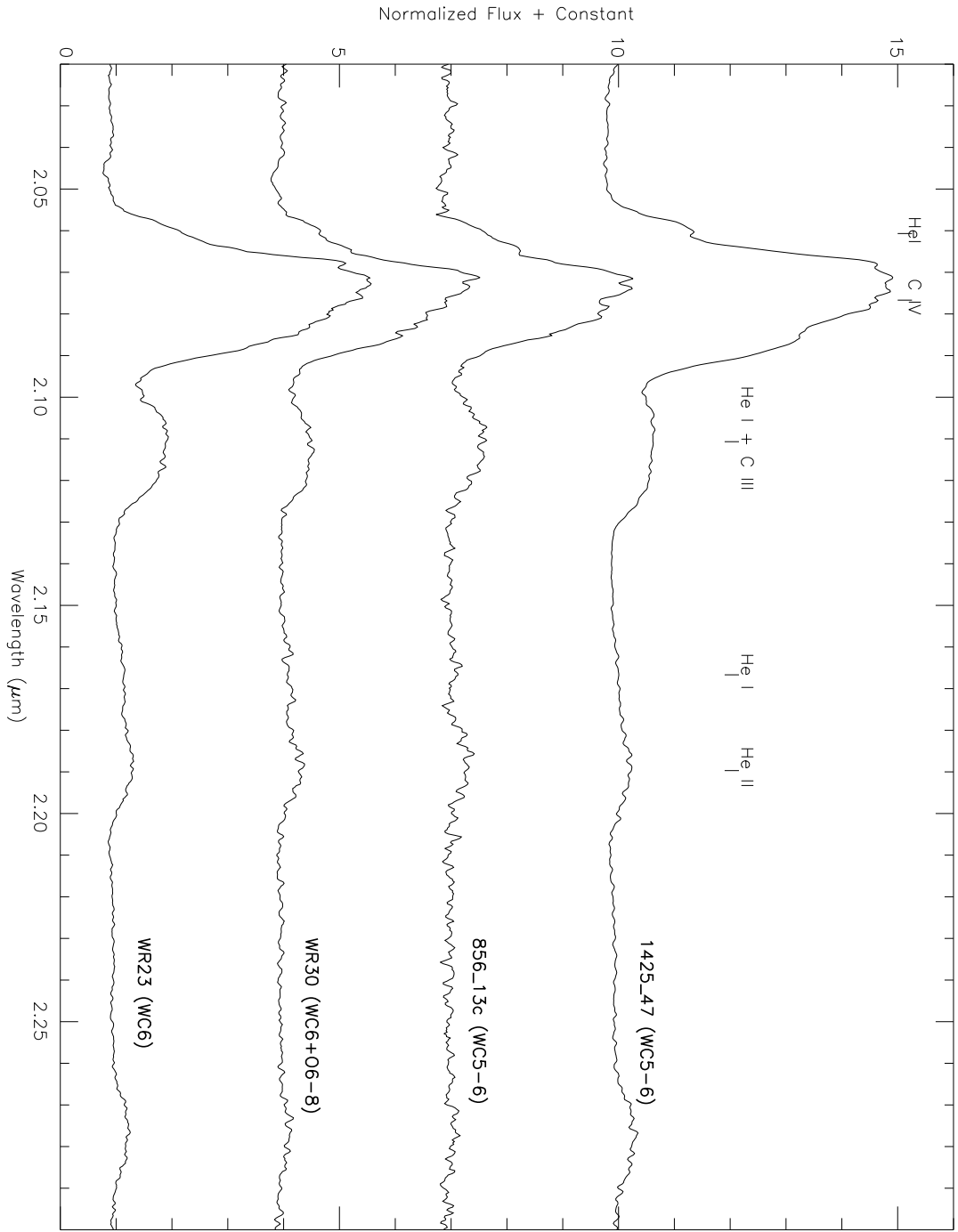}
\caption{WC5-6 spectra. \label{wc561}}
\end{figure}

\begin{figure}[t]
\figurenum{1g}
\plotone{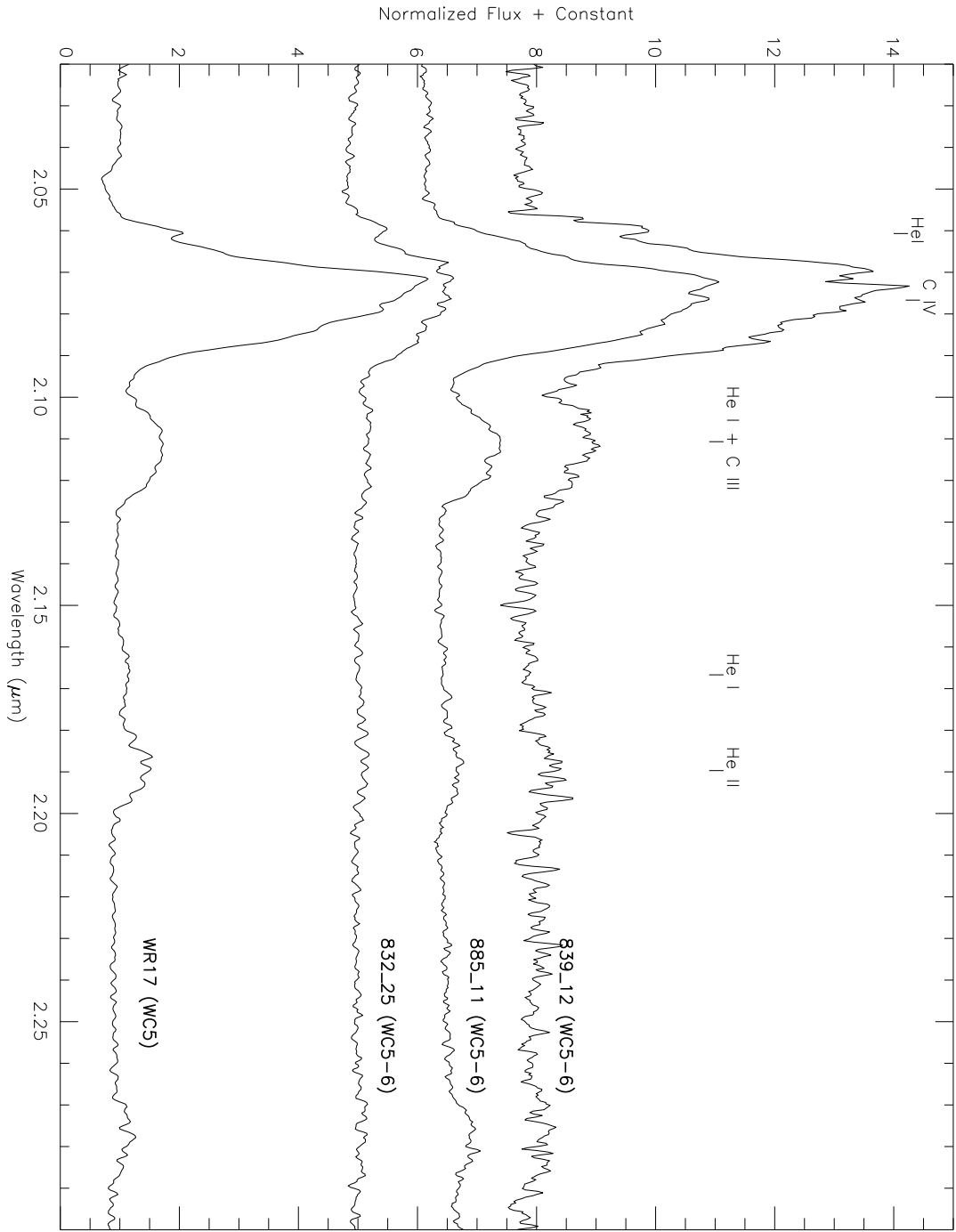}
\caption{More WC5-6 spectra. \label{wc562}}
\end{figure}

\begin{figure}[t]
\figurenum{1h}
\plotone{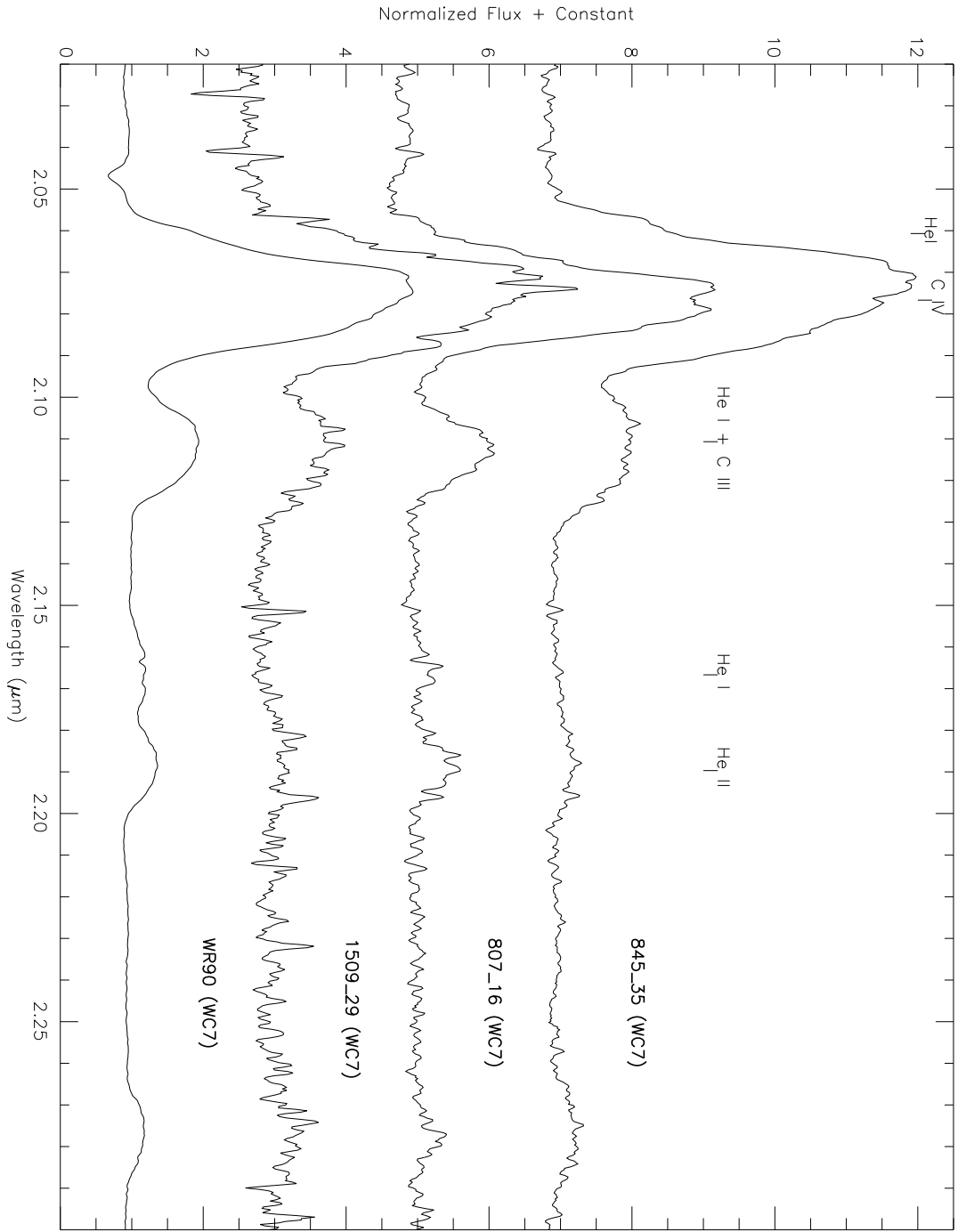}
\caption{WC7 spectra. \label{wc7}}
\end{figure}

\begin{figure}[t]
\figurenum{1i}
\plotone{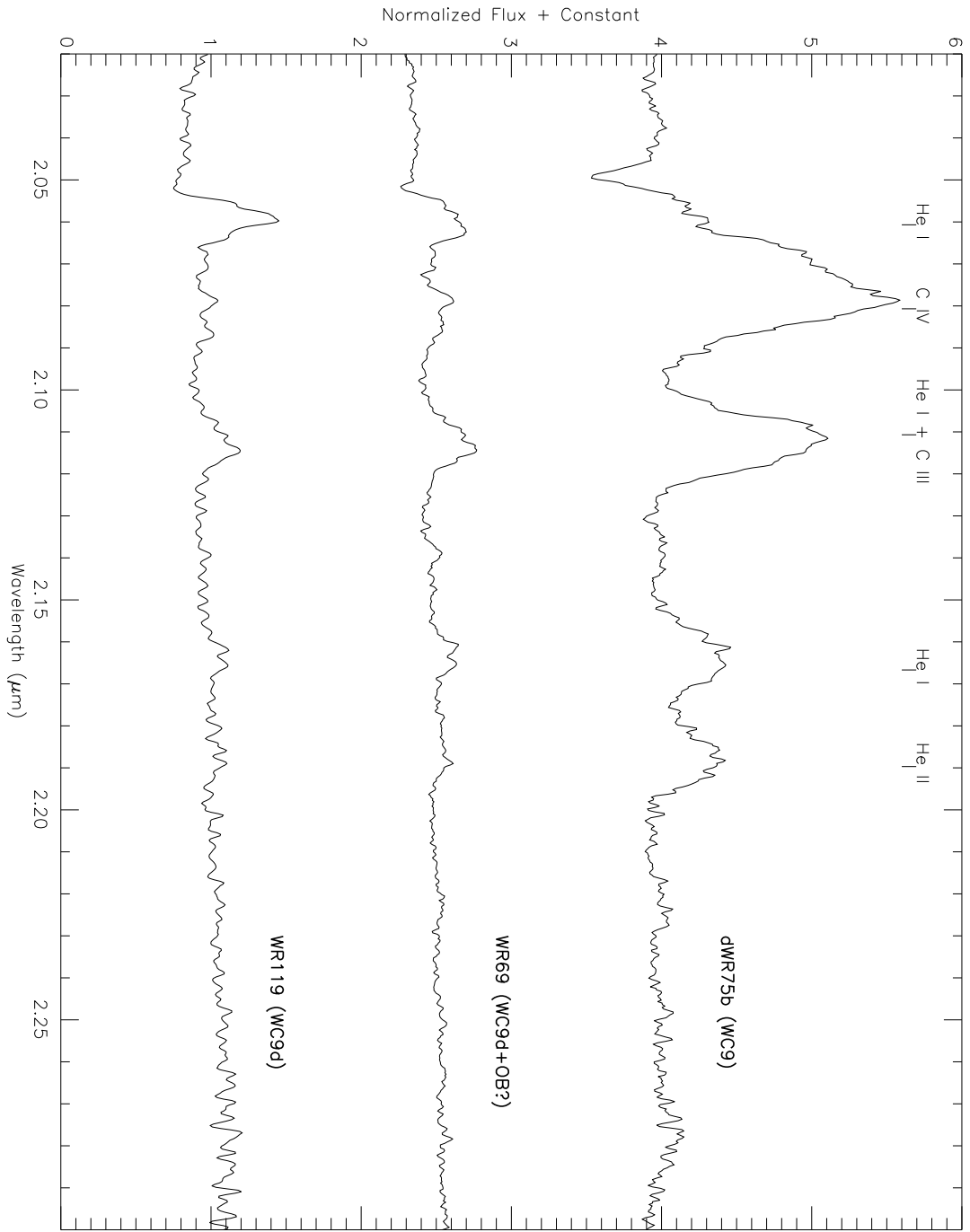}
\caption{WC9 spectra. \label{wc9}}
\end{figure}

\begin{figure}[t]
\figurenum{1j}
\plotone{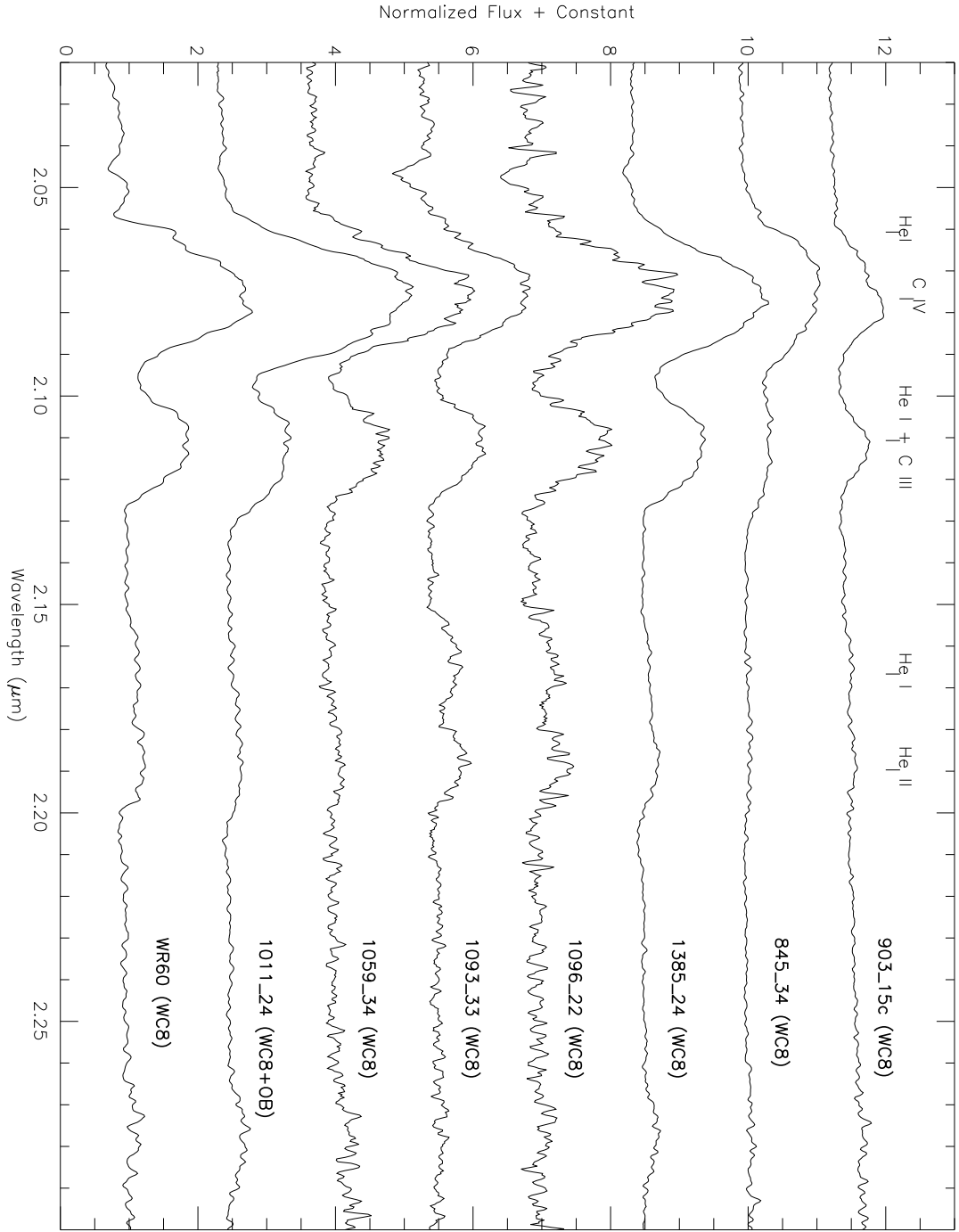}
\caption{WC8 spectra. \label{wc81}}
\end{figure}

\begin{figure}[t]
\figurenum{1k}
\plotone{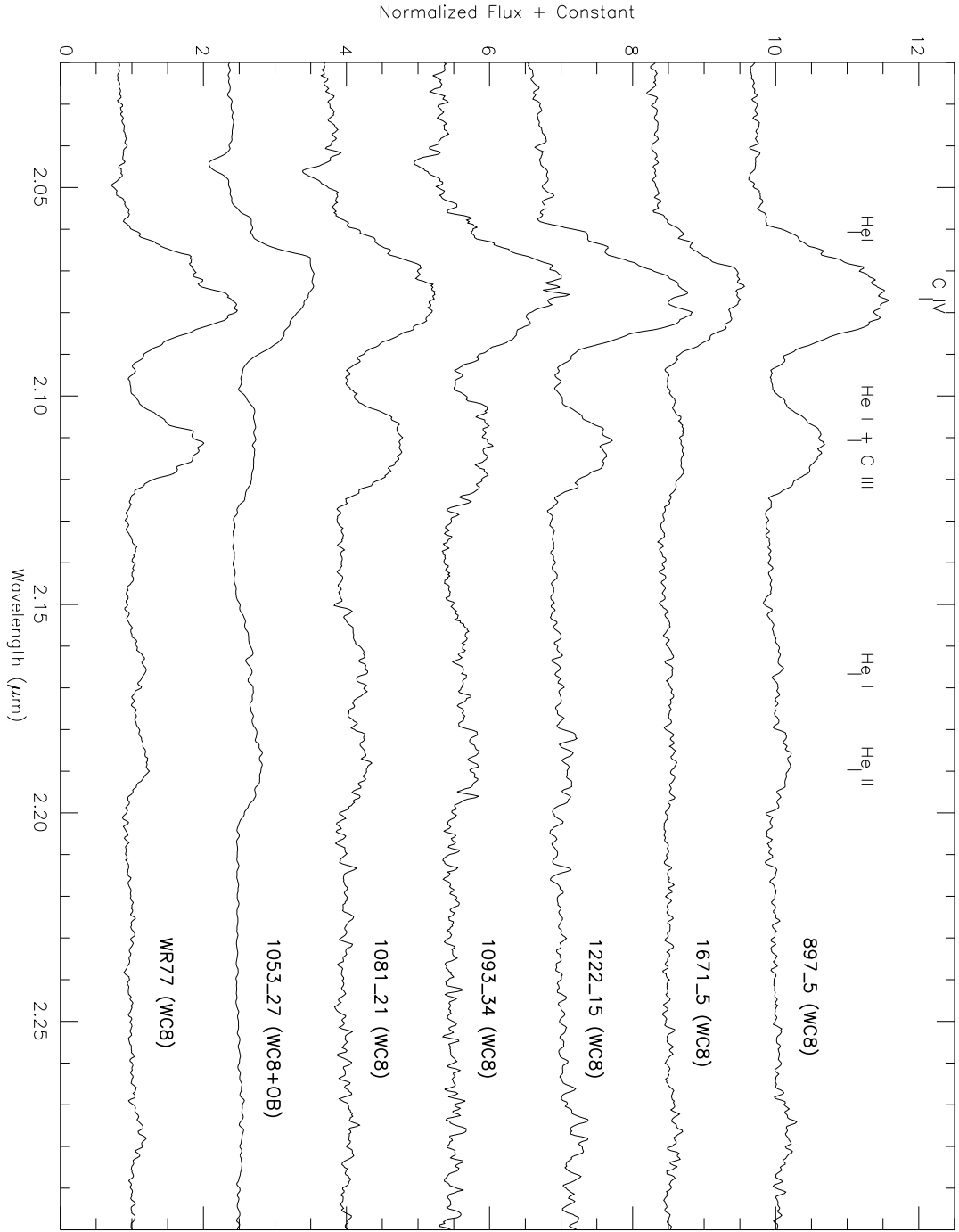}
\caption{More WC8 spectra. \label{wc82}}
\end{figure}

From our target list of 173 candidates we have discovered 41 new Wolf-Rayet stars: 15 of type WN and 26 of type WC. Right ascension and declination, as well as J, H and K$_S$ magnitudes, were obtained from 2MASS, and are listed in Table \ref{tbl-pos}. All 2MASS objects were then referenced in NOMAD \citep{zac05} to obtain B, V, and R magnitudes, when available, which are also included in Table \ref{tbl-pos}. Spectra are grouped by assigned spectral type, and are presented in Figures \ref{wn4} through \ref{wc82}.

\subsection{Spectral Line Measurements} \label{spectra}
The \textsc{iraf} task \textsc{splot} was used to to fit the continuum and then optimize and deblend Gaussian functions to fit the observed emission lines in all spectra. This gave measurements of line centers, equivalent widths (EWs) and the full-widths-at-half-maximum (FWHMs).

A number of errors contribute to reduce the accuracy of these measurements. Blending of many lines creates the largest emission features, and introduces errors into line center measurements up to 10 \AA. It also skews the shape of the feature to be a poor fit to a Gaussian. In general, the rms error of the residuals for the fits was about 10\% of the continuum value. Additionally, the continuum was quite difficult to determine precisely due to the abundance of emission lines throughout the spectrum, especially in WC stars, resulting in errors in the equivalent width measurements. A number of our WN detections were rather faint, with peak fluxes only twice the continuum level, resulting in lower signal-to-noise ratios for these spectra. Finally, ground-based observatories must peer through the murky atmosphere, so that the strong C \textsc{iv} 2.08 $\mu$m line in WC stars falls on the edge of the equally strong atmospheric absorption feature at 2.06 $\mu$m. Division by a standard star removes most of this feature; however, changes in atmospheric conditions between object and standard-star observations result in residual features on the blue side of the emission line.

\subsection{Spectral Classification} \label{speclass}

The strong emission lines visible in the spectra, while easily
distinguished as belonging to either type WC or WN, are the result of
overlapping blends of emission lines of various elements. In the optical, WR stars are categorized by looking at ratios of equivalent widths of various nitrogen species for WN, carbon and oxygen species for WC, and He \textsc{ii} and He \textsc{i} for both. However, the heavy blending of lines present in the K band makes it more difficult to find either isolated spectral lines or distinguishing ratios of blended lines \citep{fig97}.

Spectral subtypes were assigned following \cite{cro06}. These are presented in Tables \ref{tbl-eqws-wn} and \ref{tbl-eqws-wc}, along with the ratio of equivalent widths used for categorization. The ratio $W_{2.189}/W_{2.165}$ was used to categorize the WN stars, and the ratio $W_{2.076}/W_{2.110}$ was used to categorize the WC stars. Also following \cite{cro06}, WN stars with FWHM(He \textsc{ii} 2.189) $\ge$ $130 \AA $ were classified as broad/strong, and the letter 'b' appended to the subtype designation. Subtypes thus assigned are expected to be accurate to within one subtype. The 40 known WR stars were similarly assigned subtypes, which agreed within one subtype of their published spectral classifications.

Those WC stars with especially broad, heavily blended C \textsc{iv} and C \textsc{iii} lines, as presented in Figure \ref{wce} and which match our observed spectra of WR19 (WC4) and spectra observed in \cite{fig97} of WR146(WC4), WR143(WC5) and WR150(WC5), are classified more generally as WCE.

\subsection{Measured Line Centers}
Blended lines complicate the calculations of accurate radial velocities, as there is no longer any fixed line-center with which to compare theoretical and actual wavelengths. The purest line in our K band spectra, He \textsc{ii} at 21891 \AA, allows measurement of radial velocity with respect to the motion of the sun in principal; however, this line is extremely weak or not discernable in most of the WC stars. Line-center measurements are further complicated by the difficulty of accurately fitting Gaussian curves to WR emission lines, as described in section \ref{spectra}, resulting in errors of 5-10 \AA ~in determination of peak wavelengths. This corresponds to errors on the order of 50-150 km/s. In general, the direction of motion with respect to the Sun follows the clockwise rotation of the Galaxy; however, as noted, the error on these measurements is large. Measured radial velocities (based on the measured line centers, which may be shifted, depending on details of the line formation mechanism) for WNs and, when possible, WCs, are included in Tables \ref{tbl-eqws-wn} and \ref{tbl-eqws-wc}.  
\section{WR Distribution} \label{distribution}
Distances to all new WR stars were estimated from the 2MASS J, H and Ks$_S$ color excesses, using the method of spectroscopic parallax described in \cite{cro06}. Intrinsic $J-K_S$ and $H-K_S$ colors specific to WR subtype were taken from \cite{cro06} and used to calculate E$_{J-K_S}$ and E$_{H-K_S}$. Extinction ratios taken from \cite{ind05} then allow two calculations for $A_{K_S}$:

\begin{equation}
A_{K_S} = 0.67^{+0.07}_{-0.06} E_{J-K_S}
\end{equation}
and
\begin{equation}
A_{K_S} = 1.82^{+0.30}_{-0.23} E_{H-K_S}
\end{equation}

The average of these values, $\overline{A_{K_S}}$, was used to calculate the distance modulus, taking the apparent K$_S$ magnitude from 2MASS and the subtype-specific absolute magnitude, M$_{K_S}$, from \cite{cro06}. Derived K$_S$ band extinctions and Galactocentric distances, R$_G$, are listed in Table \ref{tbl-dist}. Calculations for R$_G$ assume the IAU standard Solar Galactocentric distance R$_\sun$=8.0 kpc., and the known Galactic WR stars are on the same scale in Figure 2.  
Because these calculations are based on the inherent absolute magnitude and $J-K_S$ and $H-K_S$ colors for each subtype, the errors are highly dependent on the accuracy of these measurements. The measured scatter about the adopted color values is approximately 0.02 magnitudes (table A1 in \cite{cro06}), which is negligible in A$_{K_S}$  compared to the average scatter of 0.4 magnitudes from the adopted M$_{K_S}$ values. The redundant calculations of $A_{K_S}$ also provide some indication of the reliability of our measurement. The two values are generally in agreement to within 0.2 magnitudes, especially for the WNs, however they can differ by as much as 0.66 magnitudes for some of the WCs, indicating that more accurate subtype specific colors and absolute magnitudes are needed. These uncertainties give typical errors on the order of 25\% in our distance measurements, though they may range as high as 40\% in some cases.

The distances to new stars were constrained by the limiting observable magnitude of $K_S \sim$11.5. Using the overall average extinction $A_{K_S}=1.4$, we can calculate the typical measurable distances by subtype. As all discovered WN stars were early types, we can distinguish the faintest observed, strong and weak-lined WN as having average distances of 8.0 and 9.4 kpc, respectively, while observations of the faintest WC stars yield typical distances of 8.9 kpc.

\begin{figure}[t]
\figurenum{2}
\epsscale{0.85}
\plotone{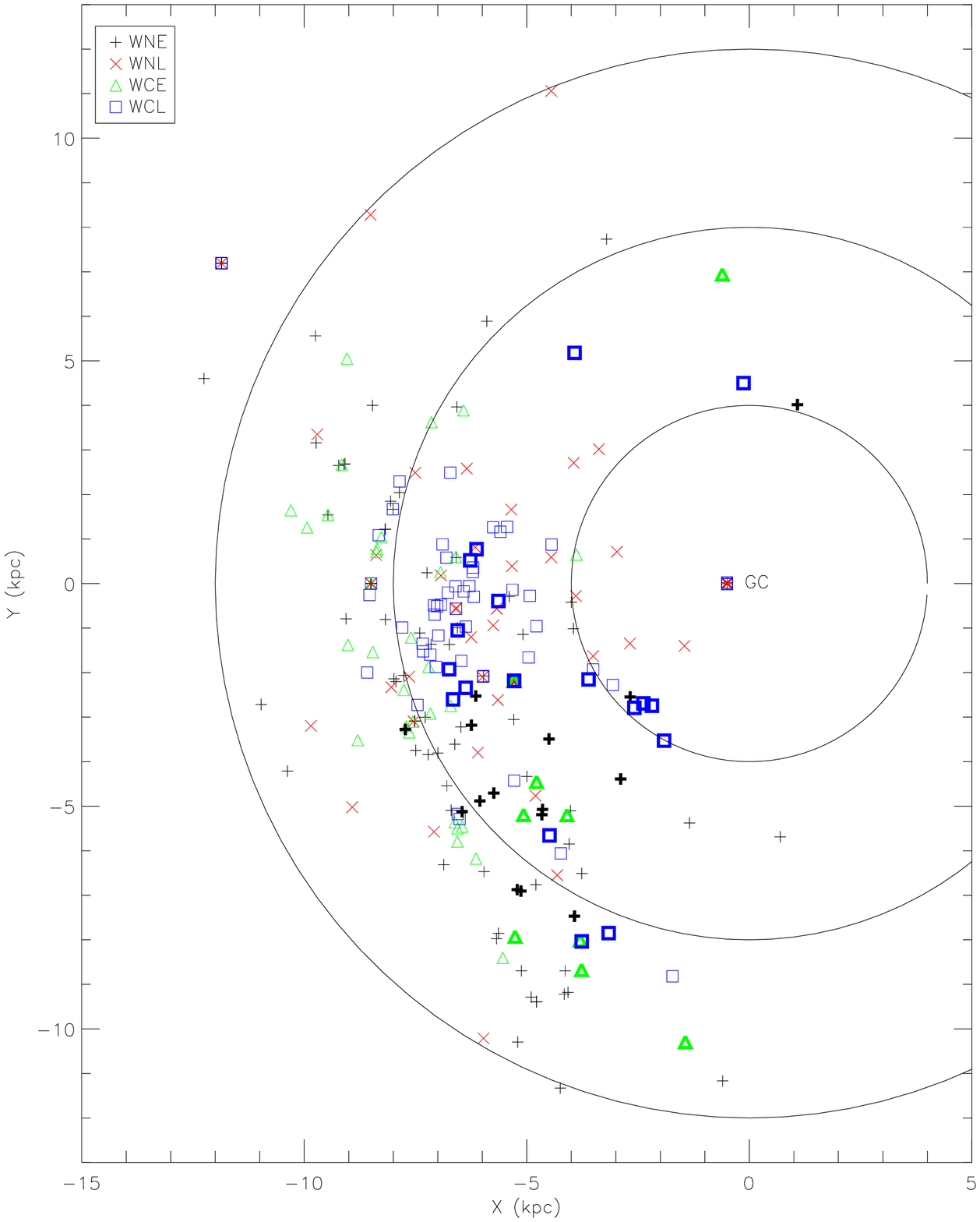}
\caption{Galactic distribution of known WR stars with estimated distances, projected on the plane. New stars are represented by bold symbols. \label{galaxyplot}}
\end{figure}

In Figure \ref{galaxyplot} our new WR stars (in bold) have been over-plotted with the previously-known WRs onto the plane of the Galaxy, with distances to the known stars taken from the 7th catalogue of Galactic Wolf-Rayet stars \citep{vdh01}. The Galactic center is labeled, and circles of radius 4, 8, and 12 kpc are plotted. The new WR stars largely follow the distribution pattern established by the known stars, though we can see that we are beginning to push out to larger heliocentric distances. We also find new stars without optical counterparts within a few kpc of our Sun, reinforcing the necessity of WR surveys in the near infrared. \cite{con90}, along with the more recent reanalysis in \cite{had07}, maintain that WR stars trace the spiral structure of the galaxy. One arm may be seen along roughly the 8 kpc radius, and an inner arm can perhaps begin to be traced along the inner 4kpc radius. However, the distance error bars are not trivial, so that firm conclusions about the utility of WR stars as spiral tracers should not yet be drawn.

\section{Conclusions}\label{conclusion}
We have discovered 41 new Galactic WR stars, 15 of type WN and 26 of type WC, using a new, near-infrared narrow-band survey of the Galactic plane. The reduced extinction from dust and gas in the near infrared makes this the optimal method for future discovery of the thousands of undetected Galactic WR stars. Of the 254 total candidates observed spectrographically, 75 proved to be emission line objects. All of the emission line objects that were not WR stars (34 objects) were planetary nebulae (PN). As the key goal of this survey is the detection of new Galactic WR stars, we amended our selection criteria to eliminate likely PN. Our modified selection criteria yielded 173 WR star candidates which were observed spectrographically: 41 proved to be new WR stars. With such a 23\% detection rate, we have barely scratched the surface of the wealth of new WR stars expected to be discovered within our survey area with the available data.

An initially fairly simple sky-subtraction methodology resulted in relatively scattered color-magnitude diagrams, raising our cut for emission objects to 5$\sigma$. It also meant that most of our non-detections were erroneously selected objects with featureless spectra. Improved sky subtraction (using entire nights of data, median-filtered in each filter as skyflats) will allow us to lower this limit to 3$\sigma$ and will improve the detection rate of emission-line objects. We expect this survey to yield thousands of additional discoveries in the coming years.

Our survey limits will be pushed fainter by the use of a larger infrared telescope for spectroscopic follow-up. As we increase the number of known stars, we will also increase the statistical significance of distribution plots, and subtype abundances, allowing us to learn more about our Galaxy's structure and composition. The Galactic center is expected to prove an especially rich area for discovery, but it is still largely terra incognita as the crowding of stars there is very high. The vast majority of Galactic Wolf-Rayet stars remain to be discovered, but we now have a proven technique to continue the search.

\acknowledgments
MMS, JG and DZ acknowledge with gratitude Hilary Lipsitz, whose ongoing support has been essential to the success of this program. AFJM, RD, LD are grateful to NSERC (Canada) and FQRNT (Quebec) for financial aid. We also thank the American Museum of Natural History for essential funding, and a careful referee for excellent suggestions. Most of the infrared imaging was expertly carried out by Claudio Aguilera, Alberto Miranda and Alberto Pasten.

\appendix
\begin{center}
{\bf APPENDIX A}
\end{center}
\section{Summary of known WRs contained in Survey Images} \label{earlywork}
In Tables 6 and 7 we give a summary of the emission line $\Delta m$ values measured for the known WRs found in the survey area. All survey stars were compared with finder charts for the known WR stars to ensure that photometric measurements were made for the correct star. All stars were found by eye to be present in the survey images, though some had not been identified photometrically due to nebulosity and occasional out-of-focus images.

The $\Delta m$ magnitude for each narrow-band filter gives the difference in magnitude from the interpolated continuum magnitude. Negative values of $\Delta m$ indicate emission, positive values indicate absorption for lines in the narrow-band emission-line filter, assuming there are no other problems. Several fields were imaged multiple times, and on more than one night. This allowed us to estimate the accuracy of the emission line magnitudes in Tables 6 and 7 as $\pm$0.03 magnitudes.

\appendix
\begin{center}
{\bf APPENDIX B}
\end{center}
\section{The Total Number and Distribution of Galactic WR Stars} \label{earlywork}

Following Shara et al. (1999), we assume that the star distribution expressed in cylindrical galactic coordinates (i.e. $N(r,l,z)$) follows that of the interstellar dust and adopt an axisymmetric exponential disk formulation with an outwards flair taken from Drimmel \& Spergel (2001):

\begin{displaymath}
\begin{array}{l}
N\left( {r,l,z} \right) = N_o e^{\frac{{ - \left( {R\left( {r,l}
\right) - R_o } \right)}}{{\alpha _{R_D } }}} \sec h^2 \left( {\frac{{z
- z_o }}{{\alpha _{H_D } \left( R \right)}}} \right) \\
\end{array}
\end{displaymath}

\noindent
With, \\
\begin{displaymath}
\begin{array}{l}
\,R\left( {r,l} \right) = \sqrt {r^2  + R_o^2  - 2rR_o \cos l}  \\
\end{array}
\end{displaymath}

\noindent
A linear flair is added by imposing the following functional form to $\alpha_{H_{D}}$:

\begin{displaymath}
\alpha_{H_{D}}(R) = \left[
\begin{array}{cc}
h_o  + (R(r,l) - r_f)h_1 & R(r,l) > r_f \\[10pt]
h_o  &  R(r,l) \leq r_f
\end{array}  \right.
\end{displaymath}

\noindent
In these equations, $N_{o}$ is the local stellar density, $R_{o} = 8000$ pc and $z_{o} = -17$ pc are respectively the solar galactocentric distance and distance from the galactic plane. The scale length $\alpha{}_{R_{D}}$ and the constants $h_{o}$, $h_{1}$ and $r_{f}$ which define the scale height $\alpha_{H_{D}}$ are taken from the fit to the FIR emission of the interstellar dust of Drimmel \& Spergel (2001) and are respectively $2260$ pc, $134$ pc, $0.015$ and $4400$ pc.

Although the assumption that the star distribution follows the dust is justified for O stars, it must be modified for WR stars to account for the observed metallicity dependence of their galactic distribution. Using the results of Maeder \& Meynet (1994) for the metallicity dependence of the WR/O number ratio and the radial metallicity distribution of the galactic disk from Smartt \& Rolleston (1997), Shara et al. (1999) found that the WR star density can be expressed as:

\begin{displaymath}
\begin{array}{l}
N_{WR} \left( {r,l,z} \right) = N_{WR_o } e^{\frac{{ - \left(
{R\left( {r,l} \right) - R_o } \right)}}{{\alpha _{R_{WR} } }}} \sec
h^2 \left( {\frac{{z - z_o }}{{\alpha _{H_D } \left( R \right)}}}
\right) \\
\end{array}
\end{displaymath}

\noindent
Where,\\

\begin{displaymath}
\begin{array}{l}
\alpha _{R_{WR} }  = \frac{1}{{\frac{1}{{\alpha _{R_D } }} + 7 \times
10^{ - 5} \ln 10}} = 1657pc, \\
\end{array}
\end{displaymath}

\noindent
Knowing the local surface density of WR stars from van der Hucht (2001):

\begin{displaymath}
\begin{array}{l}
\int\limits_{ - \infty }^\infty  {N_{WR} \left(
{0,0,z} \right)dz}  = {\rm{2}}{\rm{.87\times10}}^{{\rm{ - 6}}} pc^{ - 2}
\end{array}
\end{displaymath}

\noindent
And applying this constraint to $N_{WR}(r,l,z)$ we obtain:

\begin{displaymath}
N_{WR_o }  = \frac{{2.87 \times 10^{ - 6} pc^{ - 2} }}{{\int\limits_{
- \infty }^\infty  {\sec h^2 \left( {\frac{{z - z_o }}{{\alpha _{H_D }
\left( 0 \right)}}} \right)dz} }} = {\rm{2}}{\rm{.09\times10}}^{{\rm{ - 8}}}
pc^{ - 3}
\end{displaymath}

Finally, the conversion of the WR star density from a spatial ($N_{WR}(r,l,z)$) to a magnitude ($\eta_{WR}(m,l,z)$) dependence is done in two steps. First, we use the conservation of stars in both systems to write:

\begin{displaymath}
\eta _{WR} dmdldz = N_{WR} rdrdldz
\end{displaymath}

\noindent
Then, the radial distance $r$ is converted to the magnitude $k$ using the inverse square-law of light attenuation accounting for interstellar extinction:

\begin{displaymath}
5\log r - 5 = k - M_k  - \int_0^r {a_k \left( {r,l,z} \right)}
\,\,\,\,
\end{displaymath}

\noindent
where the extinction $a_{k}(r,l,z)$ allowing for a spherical hole in the interstellar dust at the center of the Galaxy is (see Drimmel \& Spergel 20001):

\begin{displaymath}
a_k( {r,l,z}) = \left[
\begin{array}{cc}
a_{k_o} e^\frac{-(R(r,l) - R_o)}{\alpha _{R_D }} {\rm sech}^2\left(\frac{z - z_o}{\alpha _{H_D}(R)} \right) &
R(r,l) \ge 0.5R_o  \\[15pt]
a_k(0.5R_o ,l,z)e^{- \left( \frac{R(r,l) - 0.5R_o}{2500}\right)^2}  &
R(r,l) < 0.5R_o
\end{array} \right.
\end{displaymath}

\noindent
with $a_{k_{o}} = 1.08\times10^{-4}$ mag/pc (Mathis 1990) and the intrinsic magnitude of WR stars in the $K$-band $M_{k} = -4$ mag (van der Hucht 2001).

These equations have been solved numerically to determine the star density $\eta{}_{WR}(k,l,z)$ to an accuracy better than $1\%$ using a Monte-Carlo method with a Sobol quasi-random number generator in $3D$. By integrating over the whole Galaxy the model predicts $\sim6400$ WR stars which is very close to the $6500$ predicted by van de Hucht (2001).

Further validation can be carried out by applying this model to the $V$-band where most WR stars have been detected so far. It is then possible to compare the observed to the predicted number of WR stars. To do so, we adopt $M_{V} = -5$ mag (van der Hucht 2001) and $a_{V_{o}}=1.0\times10^{-3}$ mag/pc. According to the model, the number of WR stars observable up to a magnitude of V = 15, 12 and 10 are respectively 153, 79 and 42. These numbers are very close to the respective observed numbers reported by van der Hucht (2001) of 159, 80 and 36. It is again apparent how important it is to continue the search for new WR stars in the infrared.

\begin{figure}[t]
\figurenum{3}
\epsscale{0.95}
\plotone{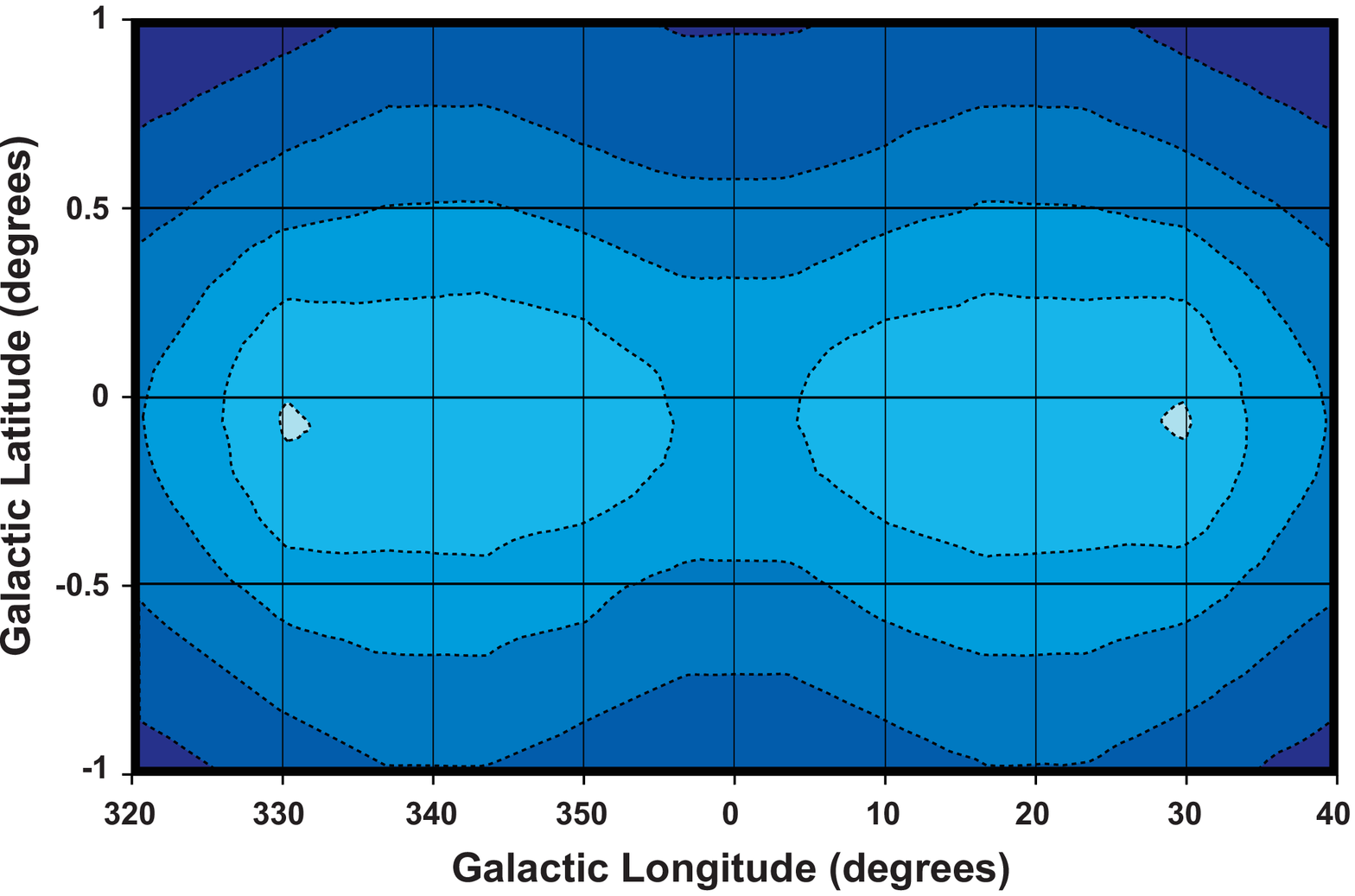}
\caption{Contour plot of the target K magnitude in order to detect 95\% of all WR stars along a line-of-sight. The inner countour represents a magnitude higher than 17 and each contour represents intervals of 1 mag. \label{con95}}
\end{figure}

\begin{figure}[t]
\figurenum{4}
\plotone{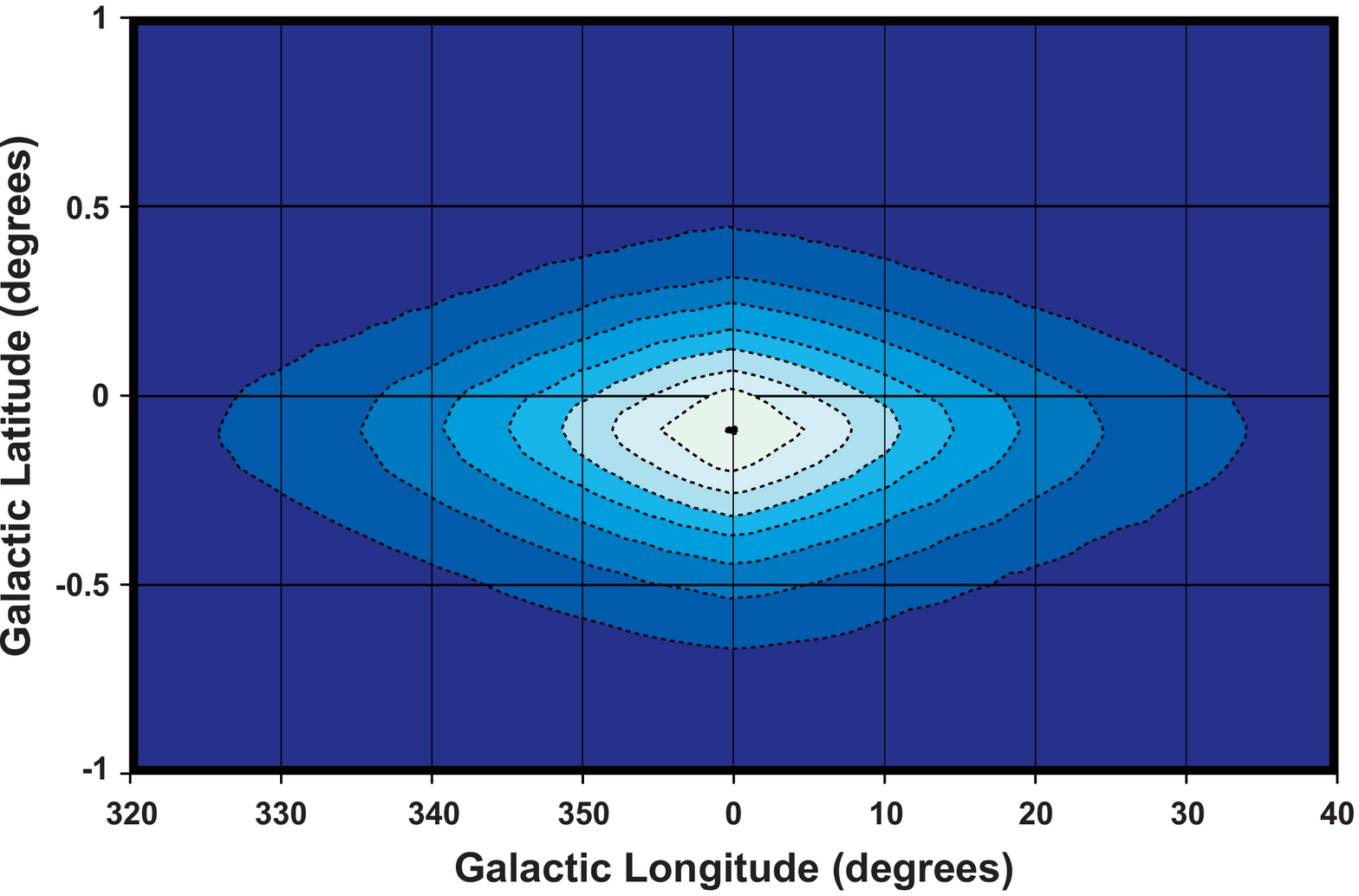}
\caption{Number of WR stars along a line-of-sight within a Cpapir field ($ 35' {\times} 35' $). The inner contour represents 40 WR stars per field and each contour represents intervals of 5 WR stars per field. According to this model, $ \sim 5600 $ of all WR stars (i.e. 88\%) should be found within the region:
$ l=320 $ to $ 400 $ and $ b = {-1} $ to $ 1 $. \label{numwr}}
\end{figure}

\begin{figure}[t]
\figurenum{5}
\plotone{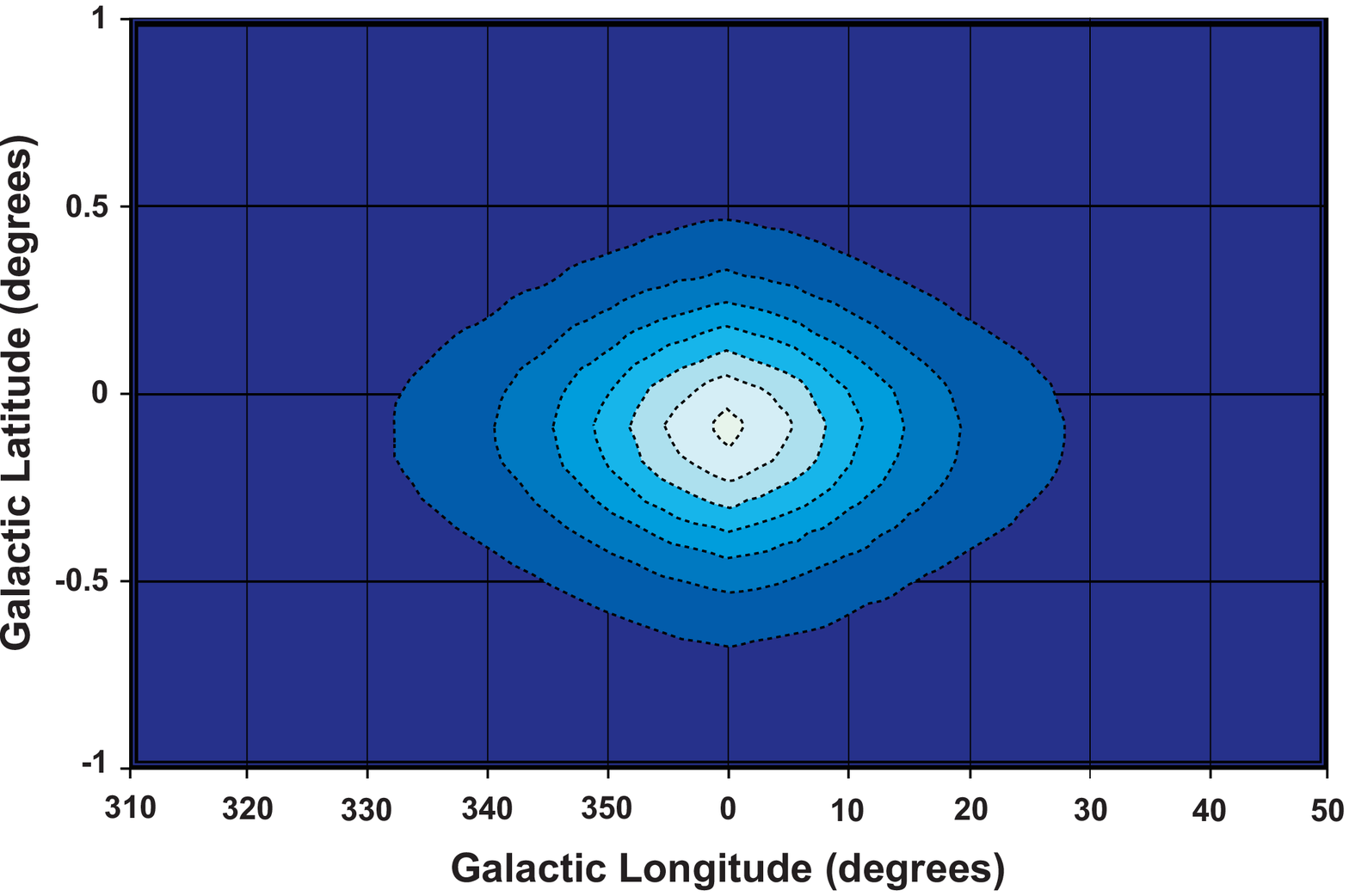}
\caption{Number of WR stars along a line-of-sight within a Cpapir field up to a magnitude K=15. The inner contour represents 35 WR stars per field and each contour represents intervals of 5 WR stars per field. \label{numwr2}}
\end{figure}

Figure \ref{con95} shows the contour plot representing the necessary target K magnitude to detect $95\%$ of all WR stars along a line of sight. Figure \ref{numwr} shows the number of WR stars expected per CPAPIR field and Figure \ref{numwr2} presents the same information for $K\leq15$. Figure 6 shows the cumulative number of expected Galactic WR stars as a function of K magnitude.
At K= 11, 12, 13 and 14, the expected numbers of Galactic WR stars are 1200, 2500, 4200 and 5400.
\appendix
\begin{center}
{\bf APPENDIX C}
\end{center}
\section{Finder Charts} \label{finders}

We present in Figures 7a through 7g the finder charts for the 41 new Wolf-Rayet stars discovered in our survey.

%\pagebreak

\begin{deluxetable}{lccc}
\tabletypesize{\tiny}
\tablecaption{The filter central wavelengths, FWHM and exposure times \label{filters}}
\tablewidth{0pt}
\tablehead{
\colhead{Filter Name} & \colhead{$\lambda$} & \colhead{$\Delta\lambda$} & \colhead{Exp-time}\\
                                      & \colhead{$\mu$m}                 & \colhead{$\mu$m}  & \colhead{seconds}
}
\startdata
CONT1 & 2.033 & 0.020 & 29.70 \\
HeI        & 2.062 & 0.010 & 59.40 \\
CIV        & 2.081 & 0.020 & 29.70 \\
Br-$\gamma$ & 2.169  &  0.020 & 29.70 \\
HeII       & 2.192 & 0.020  & 29.70 \\
CONT2 & 2.255  & 0.100 & 10.80 \\
\enddata
\end{deluxetable}

\begin{deluxetable}{l r r r r c c c r r r}
\tabletypesize{\tiny}
\tablecaption{New spectrographically confirmed WR stars. Positions and J, H, K$_S$ magnitudes were obtained from 2MASS. B, V, R magnitudes were obtained from NOMAD using their match to 2MASS.\label{tbl-pos}}
\tablewidth{0pt}
\tablehead{
\colhead{Name} &
\colhead{$\alpha$ (J2000)} & \colhead{$\delta$ (J2000)} &
\colhead{\textit{l}} & \colhead{\textit{b}} &
\colhead{B} & \colhead{V} & \colhead{R} & \colhead{J} & \colhead{H} & \colhead{K$_S$}
}
\startdata
668\_4 & 10 16 26.22 & -57 28 05.7 & 283.26 & -0.64 & - & - & 17.46 & 11.73 & 10.38 & 9.53 \\
740\_21 & 11 16 03.53 & -61 26 58.3 & 291.79 & -0.66 & 16.76 & 16.82 & 15.97 & 13.16 & 11.57 & 10.47 \\
740\_16 & 11 19 42.96 & -61 27 12.4 & 292.20 & -0.51 & 13.67 & 12.35 & 13.03 & 11.50 & 10.99 & 10.33 \\
768\_6 & 11 46 06.66 & -62 47 12.7 & 295.54 & -0.86 & 15.56 & 17.17 & 14.11 & 12.15 & 11.36 & 10.78 \\
772\_17 & 11 50 04.24 & -62 52 15.4 & 296.00 & -0.83 & 18.23 & - & 16.47 & 12.57 & 11.58 & 10.94 \\
776\_3 & 11 55 52.11 & -62 45 02.2 & 296.62 & -0.56 & 18.38 & - & 14.94 & 11.65 & 10.76 & 10.14 \\
791\_12c & 12 13 28.29 & -62 41 42.9 & 298.59 & -0.14 & - & - & - & 14.05 & 12.67 & 11.53 \\
808\_14 & 12 28 41.91 & -63 25 46.1 & 300.39 & -0.67 & 20.28 & - & 17.82 & 12.78 & 11.83 & 10.94 \\
808\_23 & 12 28 50.99 & -63 17 00.2 & 300.40 & -0.52 & - & - & 18.61 & 13.56 & 12.55 & 11.84 \\
807\_16 & 12 30 03.86 & -62 50 17.1 & 300.50 & -0.07 & - & - & 19.20 & 13.16 & 12.05 & 11.09 \\
816\_10 & 12 38 18.78 & -63 24 19.7 & 301.46 & -0.57 & - & - & 18.43 & 12.93 & 11.76 & 10.96 \\
832\_25 & 12 55 44.26 & -63 35 50.0 & 303.41 & -0.73 & - & - & - & 12.48 & 11.12 & 10.22 \\
856\_13c & 13 03 11.08 & -63 42 16.2 & 304.23 & -0.86 & - & - & 18.72 & 13.13 & 11.92 & 11.05 \\
839\_12 & 13 04 50.08 & -63 04 40.2 & 304.45 & -0.25 & - & - & - & 15.69 & 14.00 & 12.54 \\
845\_34 & 13 12 21.30 & -62 40 12.5 & 305.33 & 0.10 & - & - & 16.24 & 10.75 & 9.57 & 8.77 \\
845\_35 & 13 12 27.67 & -62 44 22.0 & 305.34 & 0.03 & - & - & 17.95 & 13.16 & 11.82 & 10.71 \\
847\_8 & 13 12 45.35 & -63 05 52.0 & 305.34 & -0.33 & - & - & - & 13.06 & 11.34 & 10.26 \\
853\_9 & 13 22 16.08 & -62 30 57.4 & 306.48 & 0.14 & - & - & - & 14.77 & 12.69 & 11.55 \\
858\_26 & 13 28 15.87 & -62 06 23.5 & 307.22 & 0.46 & - & - & 18.70 & 12.53 & 11.40 & 10.68 \\
883\_18 & 13 52 02.36 & -62 26 46.0 & 309.88 & -0.39 & - & - & - & 14.59 & 12.41 & 10.99 \\
885\_11 & 13 54 13.45 & -61 50 01.8 & 310.27 & 0.14 & - & - & - & 13.60 & 12.07 & 10.86 \\
897\_5 & 14 10 10.01 & -61 15 25.5 & 312.25 & 0.18 & - & - & - & 15.30 & 12.43 & 10.60 \\
903\_15c & 14 12 36.54 & -61 45 32.7 & 312.38 & -0.38 & - & - & - & 13.58 & 11.33 & 9.71 \\
907\_18 & 14 16 27.37 & -61 17 56.2 & 312.96 & -0.09 & - & - & - & 13.60 & 11.39 & 10.01 \\
956\_25 & 15 01 30.11 & -59 16 12.0 & 318.88 & -0.49 & - & - & - & 13.84 & 11.83 & 10.64 \\
979\_11 & 15 20 35.91 & -57 27 11.9 & 321.95 & -0.20 & - & - & - & 13.88 & 12.15 & 11.06 \\
1011\_24 & 15 43 04.68 & -55 11 12.3 & 325.81 & -0.13 & 19.93 & 17.31 & 16.06 & 10.96 & 9.94 & 9.06 \\
1053\_27 & 16 11 43.70 & -51 10 16.6 & 331.67 & 0.17 & - & - & 18.53 & 10.26 & 8.97 & 8.10 \\
1059\_34 & 16 14 37.23 & -51 26 26.3 & 331.81 & -0.34 & - & - & - & 15.05 & 12.79 & 11.54 \\
1081\_21 & 16 24 58.86 & -48 56 52.4 & 334.75 & 0.27 & - & - & - & 13.28 & 11.76 & 10.73 \\
1093\_34 & 16 31 29.23 & -47 56 16.4 & 336.22 & 0.19 & - & - & - & 15.56 & 12.86 & 11.32 \\
1093\_33 & 16 31 49.06 & -47 56 04.4 & 336.26 & 0.15 & - & - & - & 15.11 & 12.86 & 11.47 \\
1093\_53 & 16 32 12.98 & -47 50 35.8 & 336.37 & 0.17 & - & - & - & 15.00 & 12.70 & 11.34 \\
1096\_22 & 16 35 23.31 & -48 09 18.0 & 336.51 & -0.43 & - & - & - & 14.90 & 12.77 & 11.43 \\
1222\_15 & 17 22 40.74 & -35 04 52.9 & 352.20 & 0.74 & - & - & - & 15.11 & 12.05 & 10.33 \\
1385\_24 & 18 13 42.47 & -17 28 12.2 & 13.15 & 0.13 & - & - & 20.02 & 11.21 & 9.70 & 8.57 \\
1425\_47 & 18 23 03.42 & -13 10 00.4 & 18.01 & 0.18 & 15.23 & 14.49 & 14.41 & 10.34 & 9.28 & 8.27 \\
1462\_54 & 18 29 33.84 & -08 39 02.1 & 22.75 & 0.87 & - & - & - & 14.07 & 12.75 & 11.93 \\
1509\_29 & 18 41 48.45 & -04 00 12.9 & 28.27 & 0.31 & - & - & - & 15.62 & 13.32 & 11.99 \\
1613\_21 & 19 06 36.53 & +07 29 52.4 & 41.33 & 0.07 & - & - & - & 14.24 & 12.65 & 11.61 \\
1671\_5 & 19 20 40.38 & +13 50 35.2 & 48.55 & -0.05 & - & - & - & 13.57 & 11.80 & 10.76 \\
\enddata
\end{deluxetable}

\begin{deluxetable}{l r r r r r r r r r r l}
\tabletypesize{\tiny}
\tablecaption{Equivalent width (\AA) and FWHM (\AA) measurements for the most prominent lines of the new WN stars. Radial velocities (RV) are measured with respect to the He \textsc{ii} line at 21891 \AA, as it is the least blended. Uncertainties of the Ews and FWHMs are typically 20\%, while those for the RVs are discussed in the text. \label{tbl-eqws-wn}}
\tablewidth{0pt}\tablehead{
\colhead{Name} & \multicolumn{2}{c}{N \textsc{v}} & \multicolumn{2}{c}{He \textsc{i}} &
\multicolumn{2}{c}{He \textsc{ii} + Br$\gamma$} & \multicolumn{2}{c}{He \textsc{ii}} &
\colhead{$W_{2.189}/$} & \colhead{RV} & \colhead{Subtype} \\
\colhead{} & \multicolumn{2}{c}{(2.100$\mu$m)} & \multicolumn{2}{c}{(2.115$\mu$m)} &
\multicolumn{2}{c}{(2.165$\mu$m)} & \multicolumn{2}{c}{(2.189$\mu$m)} &
\colhead{$W_{2.165}$} & \colhead{(in km/s)} & \colhead{} \\
\colhead{} & \colhead{$W_{\lambda}$} & \colhead{FWHM} &
\colhead{$W_{\lambda}$} & \colhead{FWHM} &
\colhead{$W_{\lambda}$} & \colhead{FWHM} & \colhead{$W_{\lambda}$} &
\colhead{FWHM} & \colhead{} & \colhead{} & \colhead{}}
\startdata
668\_4 & -8 & 141 & -11 & 121 & -31 & 147 & -91 & 153 & 2.9 & 55 & WN5b \\
740\_21 & -14 & 208 & -20 & 200 & -39 & 165 & -203 & 242 & 5.2 & -82 & WN4b \\
768\_6 & -6 & 38 & -19 & 126 & -42 & 120 & -101 & 106 & 2.4 & 41 & WN5 \\
772\_17 & -3 & 38 & -12 & 113 & -53 & 126 & -95 & 121 & 1.8 & - & WN6 \\
776\_3 & - & - & -19 & 109 & -60 & 93 & -72 & 94 & 1.2 & - & WN6 \\
808\_23 & - & - & - & - & -46 & 101 & -141 & 126 & 3.1 & 68 & WN5 \\
816\_10 & - & - & -29 & 227 & -48 & 140 & -137 & 136 & 2.9 & -14 & WN5b \\
847\_8 & -8 & 70 & -47 & 142 & -65 & 119 & -114 & 120 & 1.8 & -96 & WN6 \\
853\_9 & - & - & -29 & 95 & -36 & 66 & -92 & 97 & 2.6 & 27 & WN6 \\
858\_26 & - & - & -26 & 96 & -38 & 74 & -56 & 78 & 1.5 & 55 & WN6 \\
907\_18 & -16 & 171 & -19 & 140 & -70 & 175 & -150 & 153 & 2.1 & -68 & WN5b \\
956\_25 & -11 & 202 & -35 & 222 & -31 & 163 & -180 & 366 & 5.8 & -315 & WN4b \\
979\_11 & - & - & -57 & 260 & -62 & 207 & -212 & 220 & 3.4 & -14 & WN4b \\
1093\_53 & - & - & -52 & 219 & -60 & 185 & -145 & 182 & 2.4 & -68 & WN5b \\
1462\_54 & - & - & -59 & 202 & -72 & 164 & -166 & 126 & 2.3 & 68 & WN5 \\
\enddata
\end{deluxetable}

\begin{deluxetable}{l r r r r r r r r r r l}
\tabletypesize{\tiny}
\tablecaption{Equivalent width (\AA) and FWHM (\AA) measurements for the most prominent lines of new WC stars. The strong C \textsc{iv} line is a blend of several emission lines. Radial velocities (RV) are measured in km/s with respect to the He \textsc{ii} line at 21891 \AA, as it is the least blended, though it is not present in all WC spectra. Uncertainties of the EWs and FWHMs are typically 20\%, while those for the RVs are discussed in the text. \label{tbl-eqws-wc}}
\tablewidth{0pt}
\tablehead{
\colhead{Name} & \multicolumn{2}{c}{C \textsc{iv}} & \multicolumn{2}{c}{He \textsc{i} + C \textsc{iii}} &
\multicolumn{2}{c}{He \textsc{i}} & \multicolumn{2}{c}{He \textsc{ii}} &
\colhead{$W_{2.076}$/} & \colhead{RV} & \colhead{Subtype} \\
\colhead{} & \multicolumn{2}{c}{(2.076$\mu$m)} & \multicolumn{2}{c}{(2.110$\mu$m)} &
\multicolumn{2}{c}{(2.165$\mu$m)} & \multicolumn{2}{c}{(2.189$\mu$m)} &
\colhead{$W_{2.110}$} & \colhead{} & \colhead{} \\
\colhead{} & \colhead{$W_{\lambda}$} & \colhead{FWHM} & \colhead{$W_{\lambda}$} & \colhead{FWHM} &
\colhead{$W_{\lambda}$} & \colhead{FWHM} & \colhead{$W_{\lambda}$} & \colhead{FWHM} &
\colhead{} & \colhead{} & \colhead{}
}
\startdata
740\_16 & -987 & 255 & -226 & 330 & - & - & -50 & 241 & 4.4 & - & WCE \\
791\_12c & -1590 & 282 & -498 & 443 & - & - & -75 & 228 & 3.2 & - & WCE \\
808\_14 & -948 & 306 & -392 & 477 & - & - & -140 & 623 & 2.4 & - & WCE \\
807\_16 & -985 & 178 & -206 & 147 & -31 & 103 & -74 & 116 & 4.8 & -55 & WC7 \\
832\_25 & -454 & 236 & -73 & 236 & -37 & 351 & -15 & 121 & 6.2 & - & WC5-6 \\
839\_12 & -1658 & 230 & -244 & 195 & - & - & -71 & 157 & 6.8 & - & WC5-6 \\
845\_34 & -374 & 271 & -131 & 284 & - & - & -35 & 384 & 2.9 & - & WC8 \\
845\_35 & -1789 & 249 & -412 & 258 & -42 & 323 & -70 & 195 & 4.3 & 0 & WC7 \\
856\_13c & -712 & 191 & -138 & 180 & -14 & 86 & -61 & 157 & 5.2 & -164 & WC5-6 \\
883\_18 & -1093 & 301 & -373 & 453 & -10 & 103 & -77 & 320 & 2.9 & - & WCE \\
885\_11 & -1489 & 210 & -279 & 180 & - & - & -60 & 174 & 5.3 & - & WC5-6 \\
897\_5 & -525 & 205 & -188 & 165 & -56 & 354 & -35 & 126 & 2.8 & - & WC8 \\
903\_15c & -165 & 186 & -66 & 123 & -13 & 125 & -14 & 125 & 2.5 & - & WC8 \\
1011\_24 & -884 & 244 & -275 & 226 & - & - & -83 & 360 & 3.2 & - & WC8 \\
1053\_27 & -303 & 222 & -81 & 213 & -59 & 223 & -84 & 202 & 3.7 & - & WC8 \\
1059\_34 & -729 & 210 & -242 & 181 & -28 & 366 & -44 & 199 & 3.0 & - & WC8 \\
1081\_21 & -419 & 219 & -214 & 174 & -75 & 185 & -60 & 143 & 2.0 & -137 & WC8 \\
1093\_33 & -369 & 199 & -155 & 161 & -69 & 161 & -69 & 132 & 2.4 & -219 & WC8 \\
1093\_34 & -475 & 226 & -178 & 222 & -62 & 212 & -84 & 187 & 2.7 & -260 & WC8 \\
1096\_22 & -571 & 202 & -230 & 156 & -71 & 169 & -76 & 141 & 2.5 & -82 & WC8 \\
1222\_15 & -551 & 197 & -171 & 154 & - & - & -45 & 230 & 3.2 & - & WC8 \\
1385\_24 & -537 & 220 & -245 & 196 & -63 & 361 & -33 & 138 & 2.2 & - & WC8 \\
1425\_47 & -1593 & 238 & -300 & 296 & -48 & 250 & -75 & 179 & 5.3 & - & WC5-6 \\
1509\_29 & -1578 & 240 & -355 & 224 & - & - & -110 & 273 & 4.4 & - & WC7 \\
1613\_21 & -659 & 365 & -245 & 518 & - & - & - & - & 2.7 & - & WCE \\
1671\_5 & -335 & 215 & -94 & 210 & -49 & 337 & -20 & 151 & 3.6 & - & WC8 \\
\enddata
\end{deluxetable}

\begin{deluxetable}{l l c c r r r r r r }
\tabletypesize{\tiny}
\tablecaption{{\footnotesize K$_S$ band extinctions and Galactocentric distances for new WR stars. Extinction, K$_S$, and distance modulus, DM, are given in magnitudes. Uncertainties in the distances are typically 25\%. Distance, d, and Galactocentric radius, R$_G$, are given in kpc.} \label{tbl-dist}}
\tablewidth{0pt}
\tablehead{
\colhead{Name} & \colhead{Subtype} &
\colhead{A$_{K_S}^{J-K_S}$} & \colhead{A$_{K_S}^{H-K_S}$} & \colhead{$\overline{A_{K_S}}$} & \colhead{K$_S$} & \colhead{M$_{K_S}$} & \colhead{DM} & \colhead{d} & \colhead{R$_G$}
}
\startdata
668\_4 & WN5b & 1.23 & 1.06 & 1.14 & 9.53 & -4.77 & 13.16 & 4.28 & 8.60\\
740\_21 & WN4b & 1.55 & 1.51 & 1.53 & 10.47 & -4.77 & 13.71 & 5.51 & 8.24\\
740\_16 & WCE & 0.37 & 0.15 & 0.26 & 10.33 & -4.59 & 14.66 & 8.56 & 9.52\\
768\_6 & WN5 & 0.80 & 0.76 & 0.78 & 10.78 & -4.41 & 14.41 & 7.62 & 8.63\\
772\_17 & WN6 & 0.97 & 0.87 & 0.92 & 10.94 & -4.41 & 14.43 & 7.68 & 8.60\\
776\_3 & WN6 & 0.89 & 0.84 & 0.86 & 10.14 & -4.41 & 13.69 & 5.46 & 7.78\\
791\_12c & WCE & 1.27 & 1.02 & 1.15 & 11.53 & -4.59 & 14.97 & 9.88 & 9.46\\
808\_14 & WCE & 0.82 & 0.56 & 0.69 & 10.94 & -4.59 & 14.84 & 9.29 & 8.87\\
808\_23 & WN5 & 1.03 & 1.00 & 1.02 & 11.84 & -4.41 & 15.23 & 11.14 & 10.02\\
807\_16 & WC7 & 0.97 & 0.69 & 0.83 & 11.09 & -4.59 & 14.85 & 9.33 & 8.88\\
816\_10 & WN5b & 1.07 & 0.96 & 1.02 & 10.96 & -4.77 & 14.71 & 8.76 & 8.44\\
832\_25 & WC5-6 & 1.10 & 0.58 & 0.84 & 10.22 & -4.59 & 13.97 & 6.22 & 7.26\\
856\_13c & WC5-6 & 0.98 & 0.53 & 0.75 & 11.05 & -4.59 & 14.89 & 9.49 & 8.46\\
839\_12 & WC5-6 & 1.70 & 1.60 & 1.65 & 12.54 & -4.59 & 15.48 & 12.48 & 10.39\\
845\_34 & WC8 & 1.04 & 0.76 & 0.90 & 8.77 & -4.65 & 12.52 & 3.19 & 7.15\\
845\_35 & WC7 & 1.23 & 0.96 & 1.10 & 10.71 & -4.59 & 14.20 & 6.93 & 7.22\\
847\_8 & WN6 & 1.76 & 1.67 & 1.71 & 10.26 & -4.41 & 12.96 & 3.90 & 7.01\\
853\_9 & WN6 & 2.04 & 1.78 & 1.91 & 11.55 & -4.41 & 14.05 & 6.46 & 6.98\\
858\_26 & WN6 & 1.12 & 1.02 & 1.07 & 10.68 & -4.41 & 14.02 & 6.37 & 6.88\\
883\_18 & WCE & 2.00 & 1.53 & 1.76 & 10.99 & -4.59 & 13.82 & 5.80 & 6.53\\
885\_11 & WC5-6 & 1.42 & 1.15 & 1.28 & 10.86 & -4.59 & 14.17 & 6.81 & 6.62\\
897\_5 & WC8 & 2.86 & 2.64 & 2.75 & 10.60 & -4.65 & 12.50 & 3.16 & 6.79\\
903\_15c & WC8 & 2.30 & 2.26 & 2.28 & 9.71 & -4.65 & 12.08 & 2.61 & 7.01\\
907\_18 & WN5b & 2.16 & 2.02 & 2.09 & 10.01 & -4.77 & 12.69 & 3.45 & 6.65\\
956\_25 & WN4b & 1.90 & 1.67 & 1.79 & 10.64 & -4.77 & 13.62 & 5.31 & 5.70\\
979\_11 & WN4b & 1.64 & 1.49 & 1.57 & 11.06 & -4.77 & 14.26 & 7.12 & 5.26\\
1011\_24 & WC8 & 0.98 & 0.91 & 0.95 & 9.06 & -4.65 & 12.76 & 3.57 & 5.90\\
1053\_27 & WC8 & 1.16 & 0.89 & 1.03 & 8.10 & -4.65 & 11.72 & 2.21 & 6.64\\
1059\_34 & WC8 & 2.06 & 1.58 & 1.82 & 11.54 & -4.65 & 14.37 & 7.47 & 4.01\\
1081\_21 & WC8 & 1.42 & 1.18 & 1.30 & 10.73 & -4.65 & 14.08 & 6.54 & 3.80\\
1093\_34 & WC8 & 2.55 & 2.11 & 2.33 & 11.32 & -4.65 & 13.64 & 5.34 & 4.21\\
1093\_33 & WC8 & 2.15 & 1.84 & 1.99 & 11.47 & -4.65 & 14.13 & 6.69 & 3.59\\
1093\_53 & WN5b & 2.20 & 1.98 & 2.09 & 11.34 & -4.77 & 14.02 & 6.36 & 3.70\\
1096\_22 & WC8 & 2.04 & 1.75 & 1.89 & 11.43 & -4.65 & 14.19 & 6.88 & 3.51\\
1222\_15 & WC8 & 2.91 & 2.44 & 2.68 & 10.33 & -4.65 & 12.30 & 2.89 & 5.65\\
1385\_24 & WC8 & 1.48 & 1.37 & 1.42 & 8.57 & -4.65 & 11.80 & 2.29 & 6.29\\
1425\_47 & WC5-6 & 0.97 & 0.78 & 0.88 & 8.27 & -4.59 & 11.98 & 2.49 & 6.18\\
1462\_54 & WN5 & 1.31 & 1.20 & 1.26 & 11.93 & -4.41 & 15.08 & 10.39 & 4.16\\
1509\_29 & WC7 & 2.02 & 1.36 & 1.69 & 11.99 & -4.59 & 14.89 & 9.50 & 4.50\\
1613\_21 & WCE & 1.35 & 0.84 & 1.09 & 11.61 & -4.59 & 15.11 & 10.51 & 6.97\\
1671\_5 & WC8 & 1.59 & 1.20 & 1.40 & 10.76 & -4.65 & 14.01 & 6.34 & 6.41\\
\enddata
\end{deluxetable}

\begin{deluxetable}{ l l r r r r }
\tabletypesize{\tiny}
\tablecaption{Known WN stars contained in survey, organized by subtype. \label{tbl-knownwns}}
\tablewidth{0pt}
\tablehead{
\colhead{WR \#} & \colhead{Subtype} & \colhead{$\Delta m_{He \textsc{i}}$} & \colhead{$\Delta m_{C \textsc{iv}}$} & \colhead{$\Delta m_{Br\gamma}$} & \colhead{$\Delta m_{He \textsc{ii}}$}
}
\startdata
48c & WN3h+WC4 & 0.01 & 0.02 & -0.37 & -0.20 \\
18 & WN4 & -0.07 & 0.06 & -0.13 & -0.32 \\
35b & WN4 & 0.10 & 0.12 & -0.18 & -0.48 \\
44a & WN4 & 0.19 & -0.22 & -0.42 & -0.19 \\
45b & WN4 & 0.11 & - & -0.19 & 0.11 \\
62a & WN4 & -0.03 & 0.01 & -0.08 & -0.13 \\
31 & WN4+O8V & -0.01 & 0.03 & -0.12 & -0.21 \\
51 & WN4+OB? & -0.08 & -0.01 & -0.15 & -0.38 \\
38a & WN5 & 0.04 & 0.02 & -0.19 & -0.25 \\
42c & WN5 & 0.09 & 0.12 & -0.16 & -0.35 \\
42d & WN5 & 0.03 & 0.06 & -0.09 & -0.29 \\
45a & WN5 & 0.08 & 0.05 & -0.17 & -0.41 \\
36 & WN5-6+OB? & 0.00 & 0.08 & -0.25 & -0.44 \\
111c & WN6 & 0.08 & 0.12 & -0.21 & -0.38 \\
35a & WN6h & 0.00 & 0.01 & -0.15 & -0.13 \\
21a & WN6+O/a & 0.09 & - & -0.05 & 0.02 \\
28 & WN6(h)+OB? & 0.06 & 0.06 & -0.21 & -0.22 \\
47 & WN6+O5V & -0.10 & 0.01 & -0.18 & -0.19 \\
55 & WN7 & -0.08 & 0.03 & -0.23 & -0.17 \\
74 & WN7 & -0.06 & 0.00 & -0.17 & -0.16 \\
84 & WN7 & -0.07 & 0.03 & -0.16 & -0.22 \\
111d & WN7? & 0.03 & 0.09 & -0.04 & -0.09 \\
19a & WN7:(h) & -0.06 & 0.00 & -0.20 & -0.10 \\
26 & WN7/WCE & 0.03 & -0.59 & -0.16 & -0.50 \\
107 & WN8 & -0.23 & -0.07 & -0.22 & -0.02 \\
%89 & WN8h+OB & - & - & - & - \\
47b & WN9h & 0.08 & 0.03 & -0.26 & -0.06 \\
108 & WN9h+OB & -0.11 & -0.02 & -0.01 & 0.12 \\
\enddata
\end{deluxetable}

\begin{deluxetable}{ l l r r r r }
\tabletypesize{\tiny}
\tablecaption{Known WC stars in survey, organized by subtype. \label{tbl-knownwcs}}
\tablewidth{0pt}
\tablehead{
\colhead{WR \#} & \colhead{Subtype} & \colhead{$\Delta m_{He \textsc{i}}$} & \colhead{$\Delta m_{C \textsc{iv}}$} & \colhead{$\Delta m_{Br\gamma}$} & \colhead{$\Delta m_{He \textsc{ii}}$}
}
\startdata
38 & WC4 & -1.06 & -1.56 & - & -0.20 \\
47c & WC5 & -0.16 & -0.42 & -0.05 & -0.08 \\
32 & WC5+OB? & -0.41 & -1.77 & -0.13 & -0.37 \\
41 & WC5+OB? & -1.06 & -2.52 & -0.40 & -79.86 \\
23 & WC6 & -0.46 & - & -0.12 & -0.28 \\
27 & WC6+a & -0.71 & -1.36 & -0.13 & -0.31 \\
31c & WC6+OB & -1.04 & -1.51 & -0.19 & -0.28 \\
38b & WC7+OB & -0.77 & -1.14 & -0.08 & -0.11 \\
39 & WC7+OB? & -0.31 & -0.76 & -0.13 & -0.17 \\
42 & WC7+O7V & -0.36 & - & -0.11 & -0.12 \\
50 & WC7+OB & -0.76 & -1.42 & -0.02 & -0.15 \\
111b & WC9d & 0.12 & 0.18 & -0.11 & -0.19 \\
48b & WC9d & -0.03 & -0.06 & - & -0.24 \\
\enddata
\end{deluxetable}

\begin{figure}[t]
\figurenum{6}
\plotone{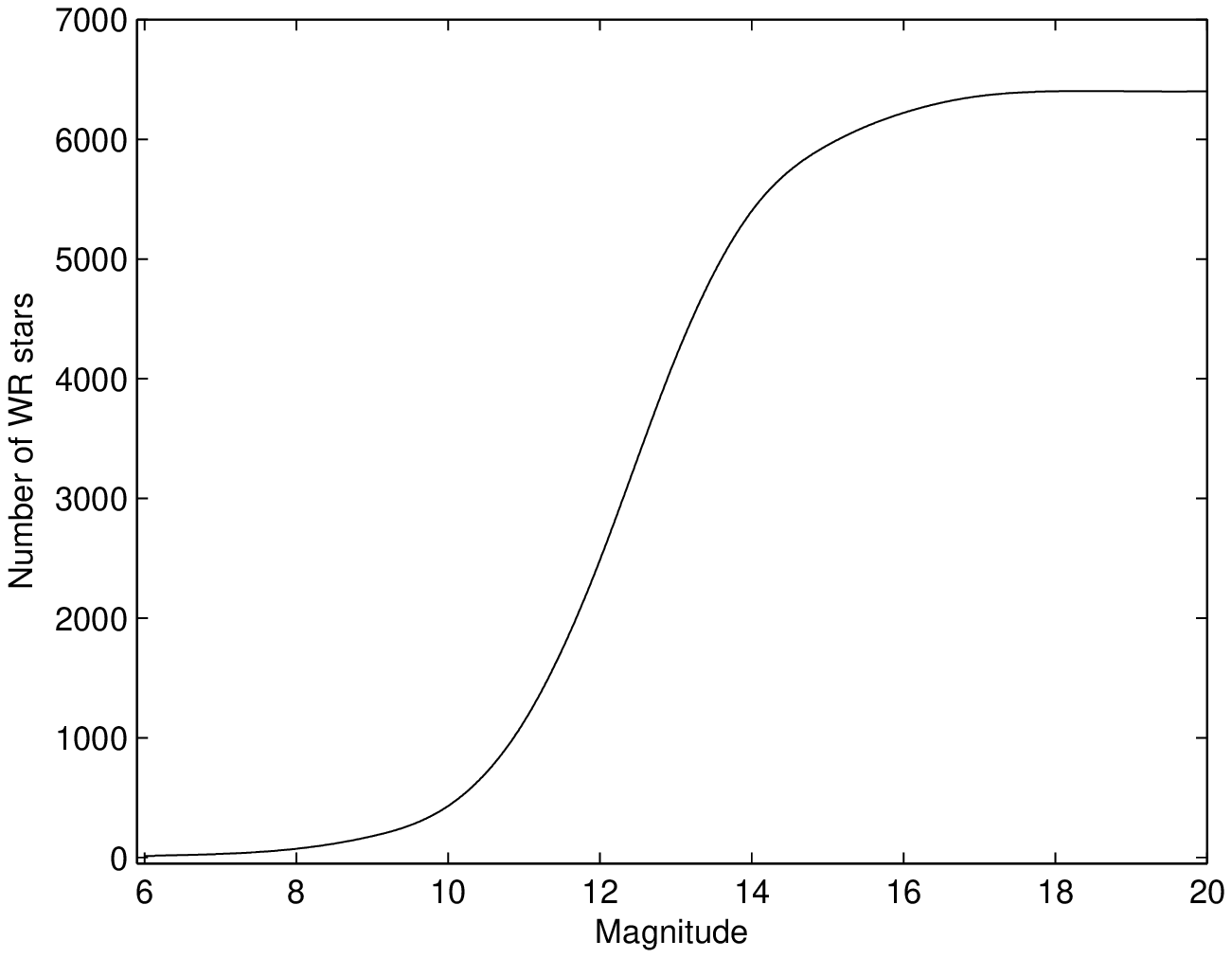}
\caption{Predicted cumulative number of WR stars as a function of K magnitude.\label{cumplot}}
\end{figure}
\clearpage

\begin{figure}[th]
\figurenum{7a}
\epsscale{0.95}
\plotone{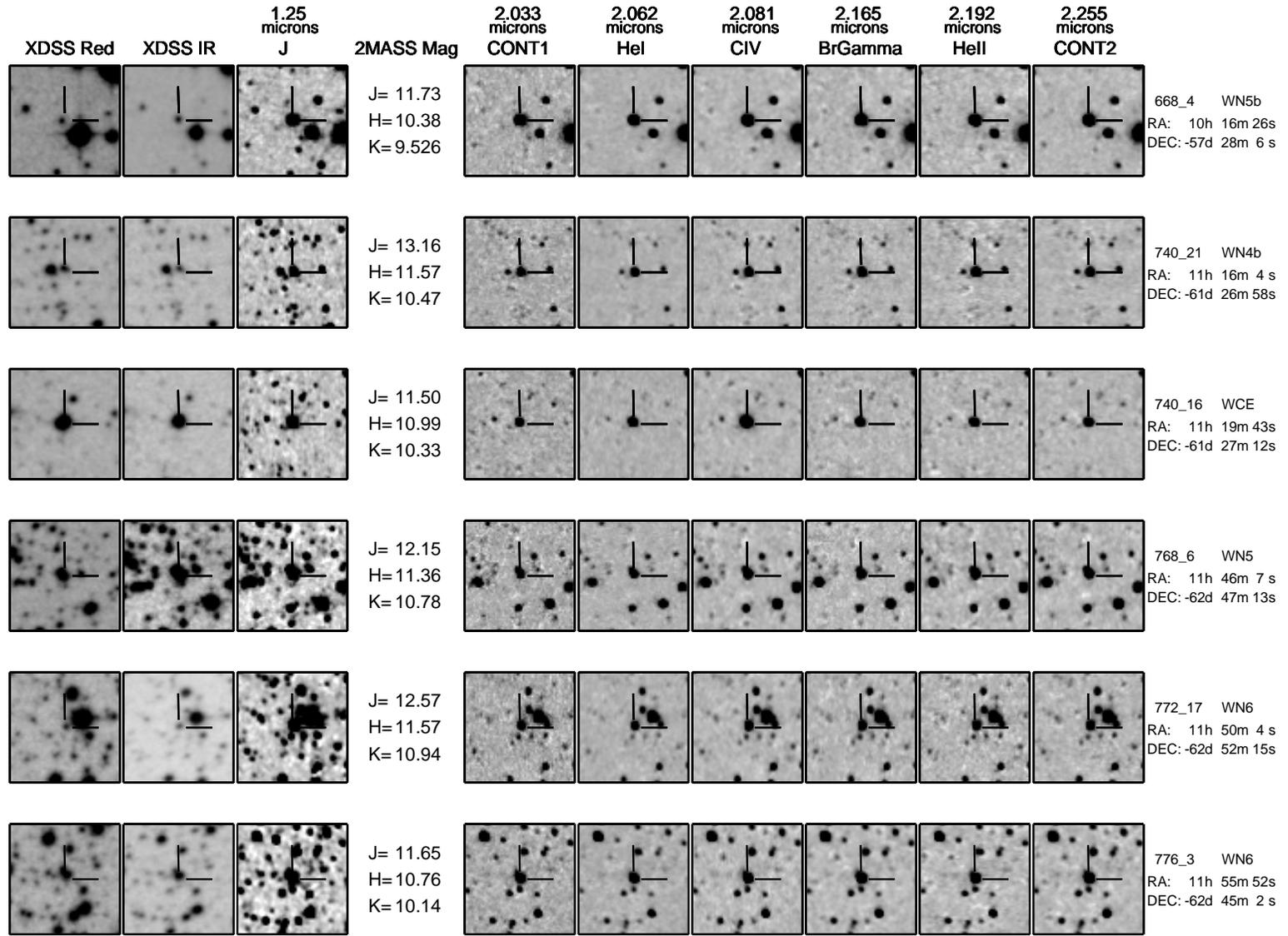}
\caption{Finder Charts for WR stars in table 2. \label{finder1} }
\end{figure}

\begin{figure}[th]
\figurenum{7b}
\plotone{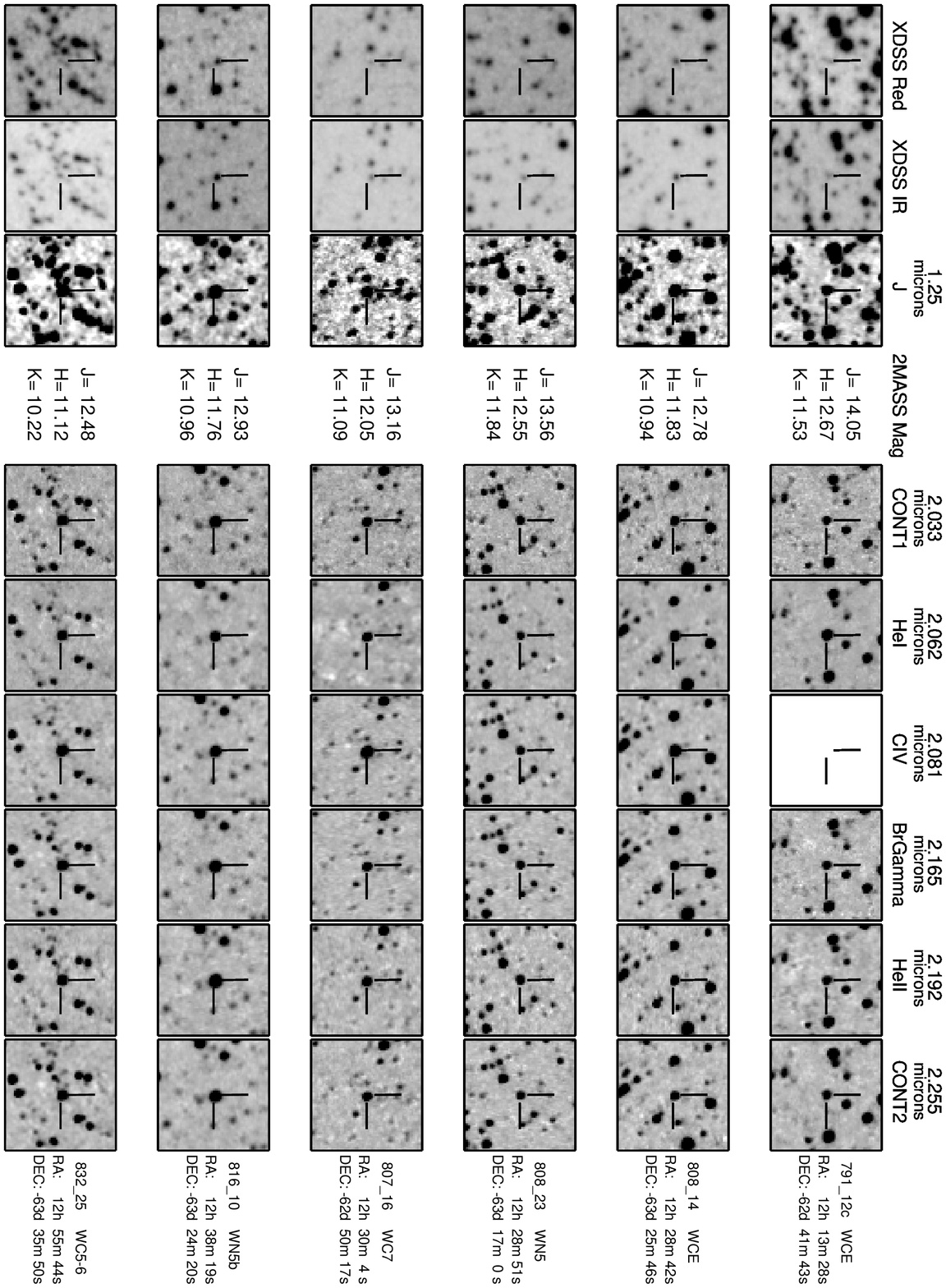}
\caption{Continued \label{finder2} }
\end{figure}

\begin{figure}[t]
\figurenum{7c}
\plotone{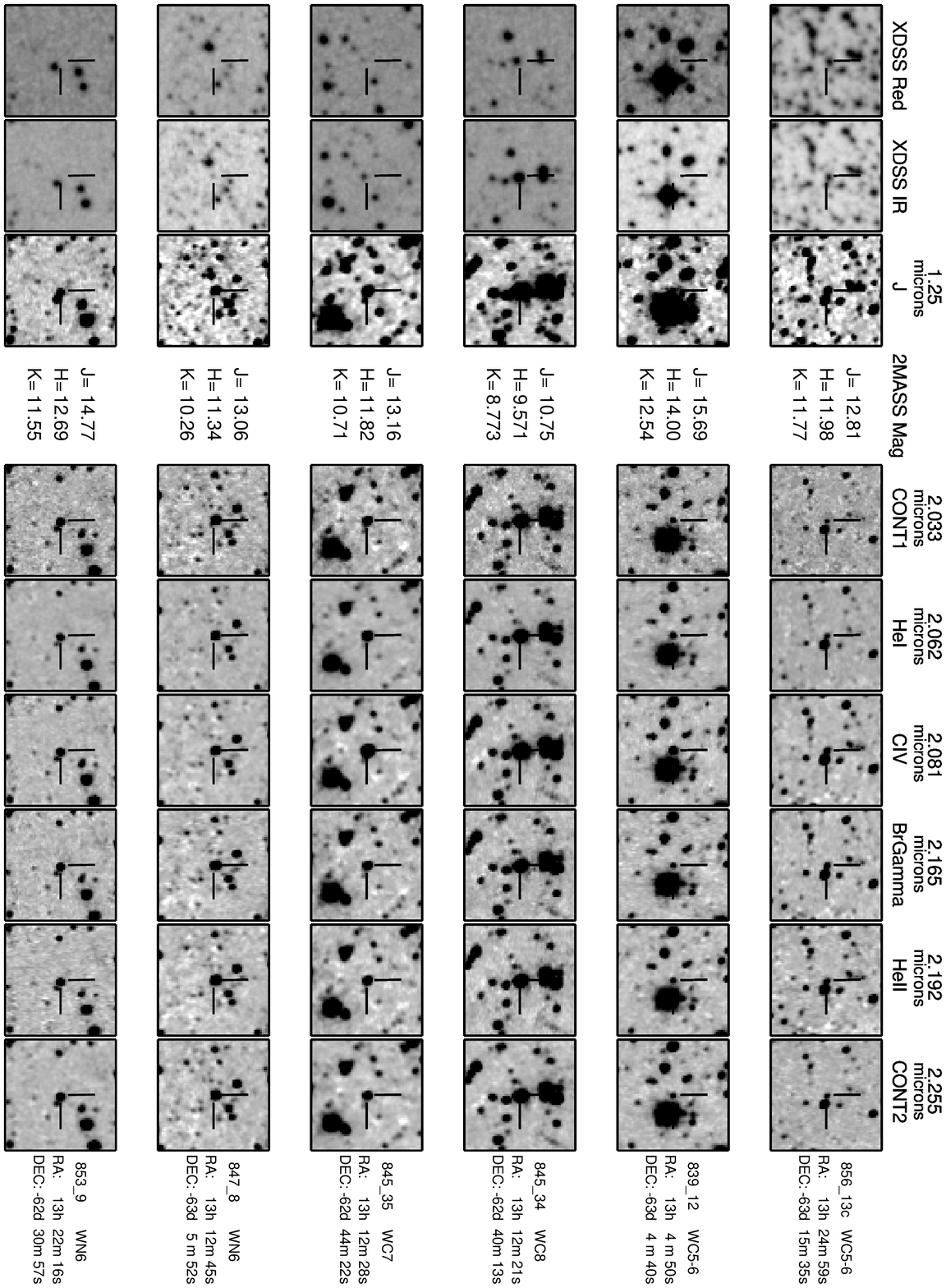}
\caption{Continued \label{finder3}}
\end{figure}

\begin{figure}[t]
\figurenum{7d}
\plotone{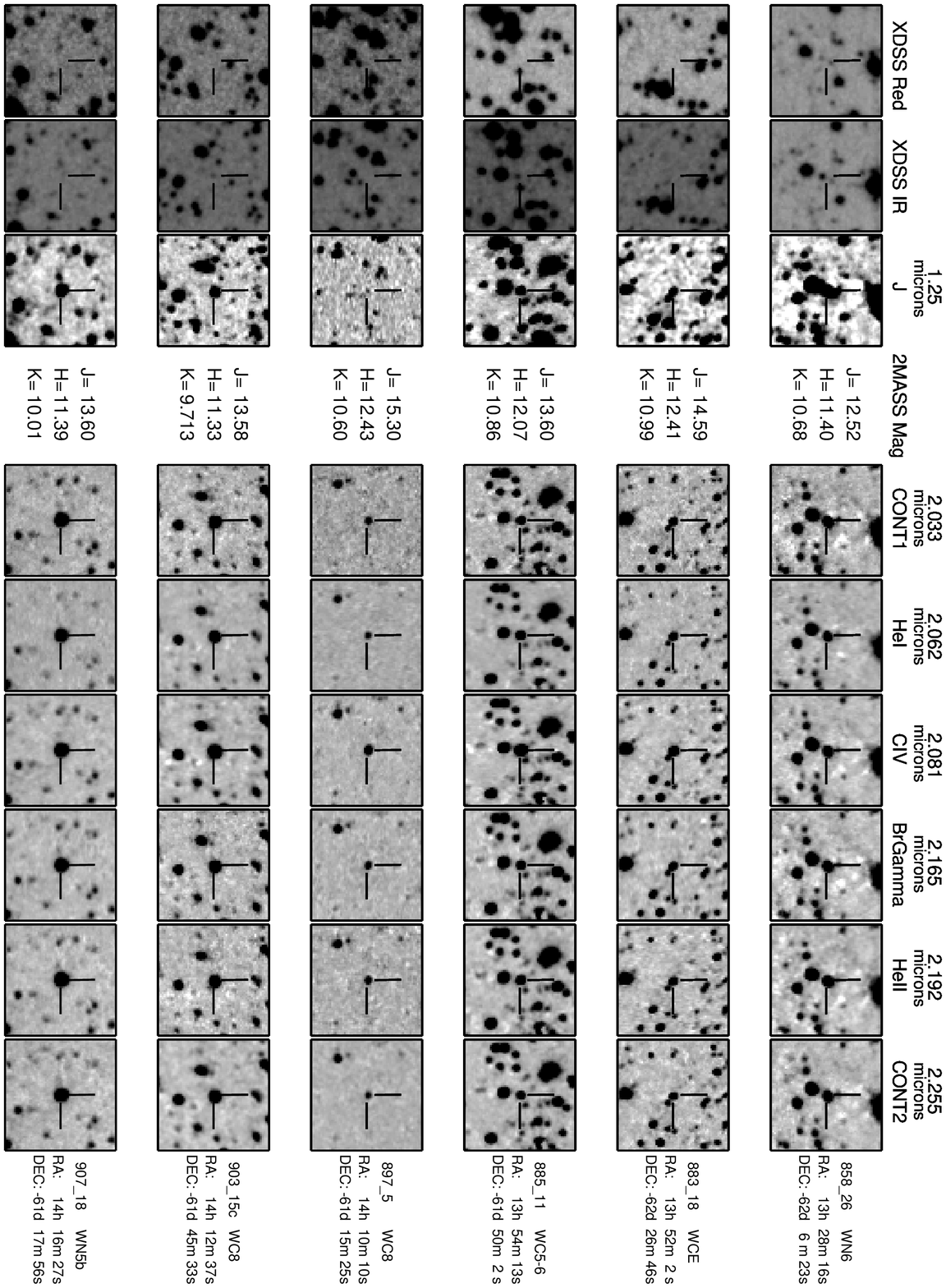}
\caption{Continued \label{finder4}}
\end{figure}

\begin{figure}[t]
\figurenum{7e}
\plotone{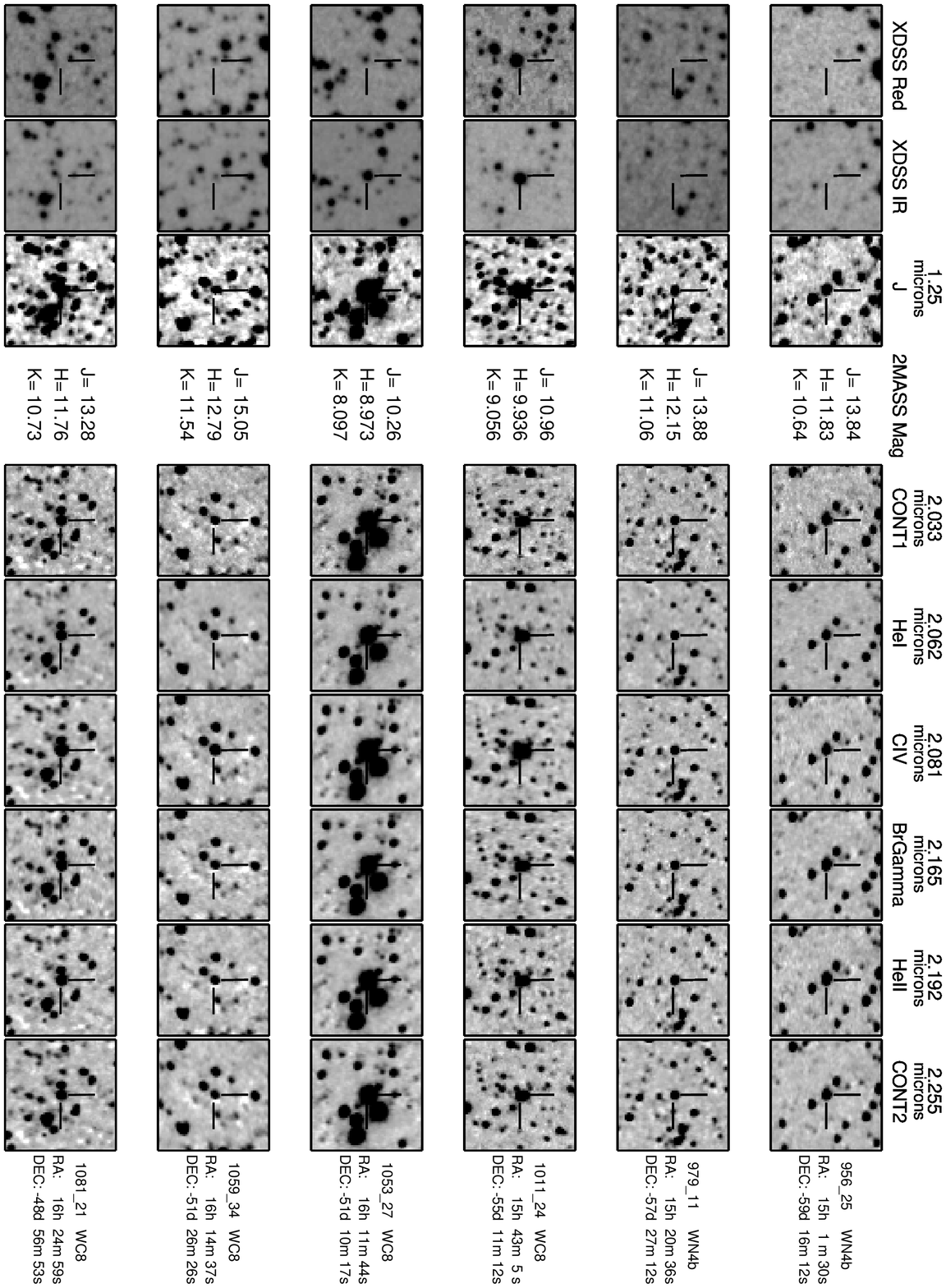}
\caption{Continued \label{finder5}}
\end{figure}

\begin{figure}[t]
\figurenum{7f}
\plotone{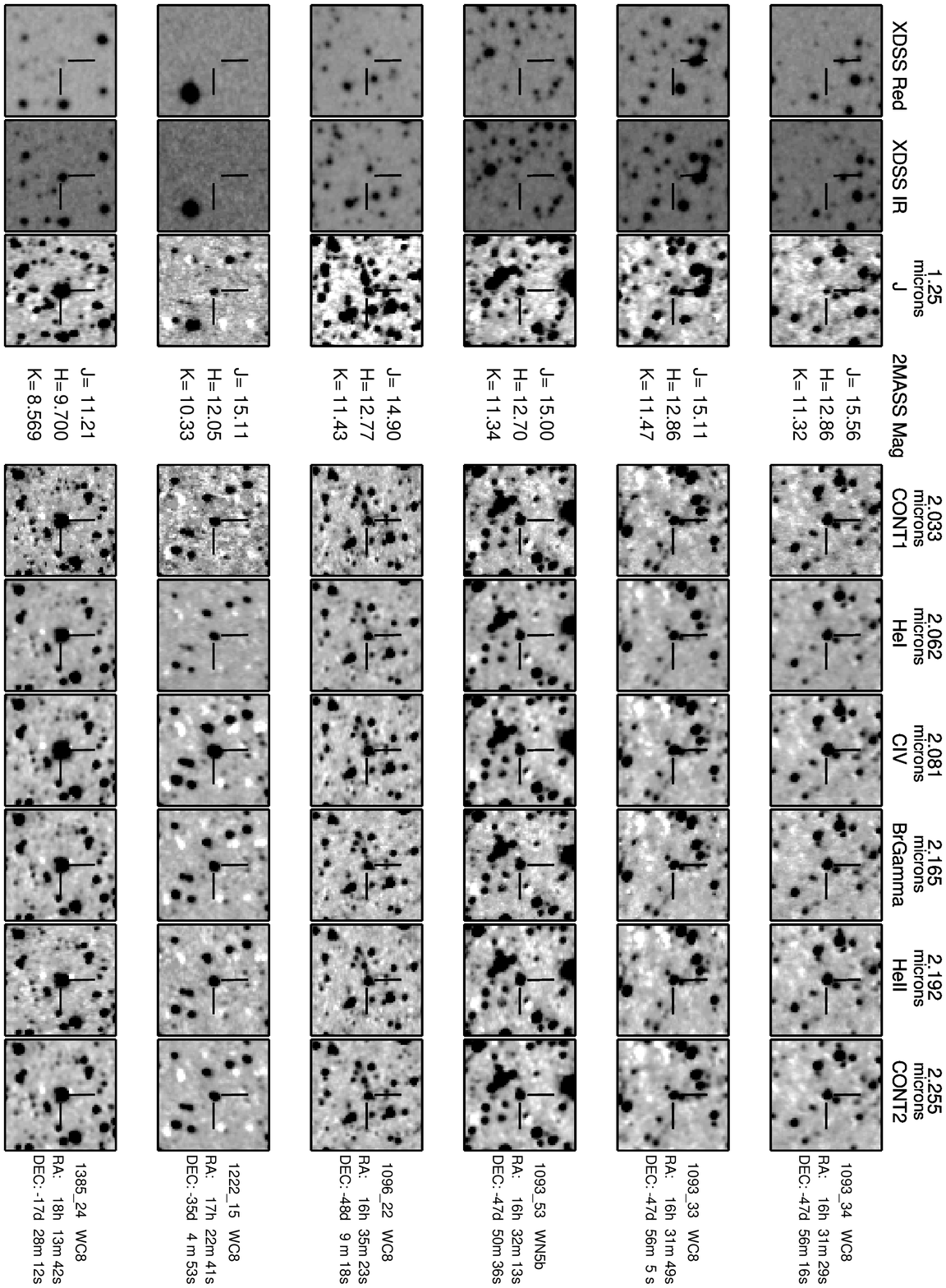}
\caption{Continued \label{finder6}}
\end{figure}

\begin{figure}[t]
\figurenum{7g}
\plotone{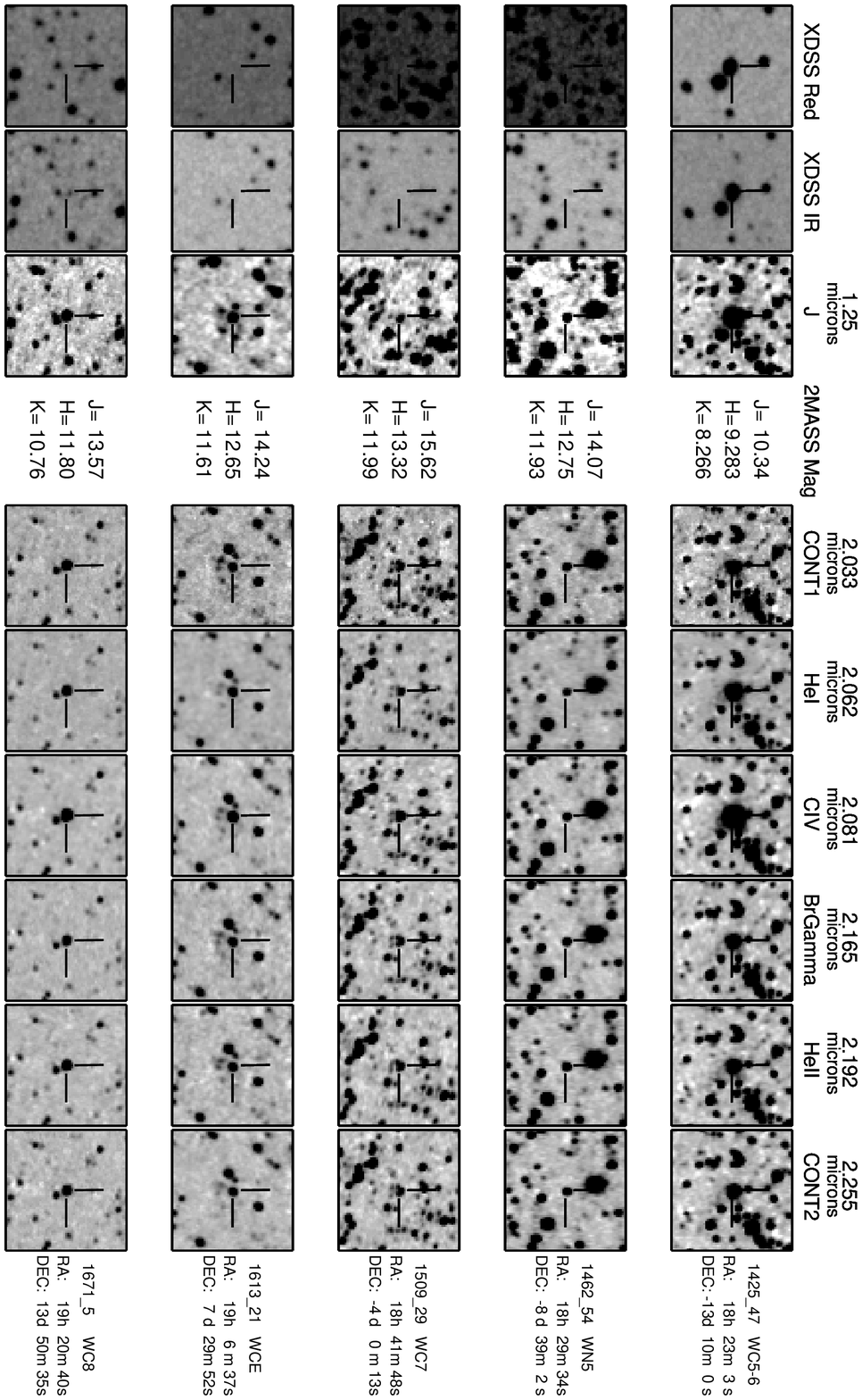}
\caption{Continued \label{finder7}}
\end{figure}

\end{document}